\documentstyle[aps,pre,epsf,cite,epsfig,rotate,multicol]{revtex}

\begin{document}
\title{Effect of Trends on Detrended Fluctuation Analysis}
\author{Kun Hu$^1$, Plamen~Ch.~Ivanov$^1$$^2$, Zhi Chen$^1$, Pedro Carpena$^3$, H.~Eugene~Stanley$^1$}

\address{ $^1$ Center for Polymer Studies and Department of Physics,
                Boston University, Boston, MA 02215\\
    $^2$ Harvard Medical School, Beth Israel Deaconess Medical Center, Boston, MA 02215\\
    $^3$ Departamento de F\'{\i}sica Aplicada II, Universidad de
                M\'alaga E-29071, Spain\\}

\maketitle
\begin{abstract}

Detrended fluctuation analysis (DFA) is a scaling analysis method used to
estimate long-range power-law correlation exponents in noisy signals.  Many
noisy signals in real systems display trends, so that the scaling results
obtained from the DFA method become difficult to analyze. We systematically
study the effects of three types of trends --- linear, periodic, and
power-law trends, and offer examples where these trends are likely to occur
in real data. We compare the difference between the scaling results for
artificially generated correlated noise and correlated noise with a trend,
and study how trends lead to the appearance of crossovers in the scaling
behavior. We find that crossovers result from the competition between the
scaling of the noise and the ``apparent'' scaling of the trend. We study how
the characteristics of these crossovers depend on (i) the slope of the linear
trend; (ii) the amplitude and period of the periodic trend; (iii) the
amplitude and power of the power-law trend and (iv) the length as well as the
correlation properties of the noise. Surprisingly, we find that the
crossovers in the scaling of noisy signals with trends also follow scaling
laws --- i.e. long-range power-law dependence of the position of the
crossover on the parameters of the trends. We show that the DFA result of
noise with a trend can be exactly determined by the superposition of the
separate results of the DFA on the noise and on the trend, assuming that the
noise and the trend are not correlated.  If this superposition rule is not
followed, this is an indication that the noise and the superimposed trend are
not independent, so that removing the trend could lead to changes in the
correlation properties of the noise. In addition, we show how to use DFA
appropriately to minimize the effects of trends, and how to recognize if a
crossover indicates indeed a transition from one type to a different type of
underlying correlation, or the crossover is due to a trend without any
transition in the dynamical properties of the noise.
\end{abstract}
\begin{multicols}{2}
\section{Introduction}
Many physical and biological systems exhibit complex behavior
characterized by long-range power-law correlations. Traditional
approaches such as the power-spectrum and correlation analysis are
not suited to accurately quantify long-range correlations in
non-stationary signals --- e.g. signals exhibiting fluctuations
along polynomial trends. Detrended fluctuation analysis
(DFA)\cite{CKDFA1,SVDFA1,SMDFA1,taqqu95} is a scaling analysis method
providing a simple quantitative parameter --- the scaling exponent
$\alpha$ --- to represent the correlation properties of a signal.
The advantages of DFA over many methods are that it permits the detection of long-range correlations embedded in seemingly non-stationary time series, and also avoids the spurious detection of apparent long-range correlations that are artifact of non-stationarity. In the past few years, more than 100 publications have
utilized the DFA as method of correlation analysis, and have
uncovered long-range power-law correlations in many research
fields such as cardiac
dynamics\cite{iyengaramjphsiolreg,plamennature1996,HOcirc1997,plamenphsa1998,barbiheartchaos1998,plamenuropl1999,Pikkujamsaheartcir1999,solomrev1999,Genephsa1999,ashkenazyheartfrac1999,makikallioheartamjcardiol1999,crossoverCK,CKfractal,Absil1999,solomphsa1999,toweillheartmed2000,bundesleep2000,Laitio2000,Yosef2001},
bioinformatics\cite{SVDFA1,CKDFA1,rmsCK,genenuovodna1994,mantegnaprl1994,ckdnapha1995,solomdnafractal1995,mantegnaprl1996,SVdnapha1998,blesicdnapha1999,Yoshinagaphya2000,Perazzofractal2000,Siwy},
economics\cite{Liu97,vandewallephsa1997,vandewallepre1998,Liu99,janosiecopha1999,ausloosphsa1999_12,robertoecopha1999,Vandewalle1999,grau-carles2000,ausloosphsa2000_9,ausloosphsa2000_10,ausloospre2001,ausloosIntJModPhys2001},
meteorology\cite{Ivanovameteo1999_12,Montanari2000,Matsoukas2000},
geology\cite{malamudjstatlaninfer1999}, ethology\cite{Alados2000} etc.
Furthermore, the DFA method may help identify different states of
the same system according to its different scaling behaviors ---
e.g. the scaling exponent $\alpha$ for heart inter-beat intervals
is different for healthy and sick
individuals\cite{ashkenazyheartfrac1999,mfCK_lett,CKfractal,crossoverCK}.

The correct interpretation of the scaling results obtained by the
DFA method is crucial for understanding the intrinsic dynamics of
the systems under study. In fact, for all systems where the DFA
method was applied, there are many issues that remain unexplained.
One of the common challenges is that the correlation exponent is
not always a constant (independent of scale) and crossovers often exist --- i.e. change
of the scaling exponent $\alpha$ for different range of
scales\cite{crossoverCK, Liu97, iyengaramjphsiolreg}. A crossover
usually can arise from a change in the correlation
properties of the signal at different time or space scales, or can often arise from trends in the data. In
this paper, we systematically study how different types of trends
affect the apparent scaling behavior of long-range correlated signals.
 The
existence of trends in times series generated by physical or
biological systems is so common that it is almost unavoidable. For
example, the number of particles emitted by a radiation source in
an unit time has a trend of decreasing because the source becomes
weaker\cite{Nradiation,gandiradiation}; the density of air due to gravity has a
trend at different altitude \cite{Bundeatm}; the
air temperature in different geographic locations and the water
flow of rivers have a periodic trend due to seasonal
changes\cite{bundetem,Ivanovacloud2000,talknertem2000,Montanari2000,Matsoukas2000}; the
occurrence rate of earthquakes in certain area has trend in
different time period\cite{ogataearth}. An immediate problem
facing researchers applying scaling analysis to time series is
whether trends in data arise from external conditions, having
little to do with the intrinsic dynamics of the system generating
noisy fluctuating data. In this case, a possible approach is to
first recognize and filter out the trends before we attempt to
quantify correlations in the noise. Alternatively, trends may
arise from the intrinsic dynamics of the system, rather than being an
epiphenomenon of external conditions, and thus may be correlated
with the noisy fluctuations generated by the system. In this case,
careful considerations should be given if trends should be
filtered out when estimating correlations in the noise, since such
"intrinsic" trends may be related to the local properties of the
noisy fluctuations.

Here we study the origin and the properties of crossovers in the
scaling behavior of noisy signals, by applying the DFA method
first on correlated noise and then on noise with trends, and
comparing the difference in the scaling results. To this end, we
generate artificial time series --- anticorrelated, white and
correlated noise with standard deviation equal to one --- using
the modified Fourier filtering method introduced by Makse et
al.\cite{MFFM}. We consider the case when the trend is independent
of the local properties of the noise (external trend). We find
that the scaling behavior of noise with a trend is a superposition
of the scaling of the noise and the apparent scaling of the trend,
and we derive analytical relations based on the DFA, which we call
``superposition rule''. We show how this ``superposition rule'' can
be used to determine if the trends are independent of the noisy
fluctuation in real data, and if filtering these trends out will
no affect the scaling properties of the data.


The outline of this paper is as follows. In Sec.\ref{secdfa}, we
review the algorithm of the DFA method, and in Appendix
\ref{secpuren} we compare the performance of the DFA with the
classical scaling analysis ---Hurst's analysis (R/S analysis)---
and show that the DFA is a superior method to quantify the scaling
behavior of noisy signals. In Sec.~\ref{seclin}, we consider the
effect of a linear trend and we present an analytic derivation of
the apparent scaling behavior of a linear trend in Appendix
\ref{secdfa1l}. In Sec.~\ref{secsin}, we study a periodic trend,
and in Sec.~\ref{seclc} the effect of power-law trend. We
systematically study all resulting crossovers, their conditions of
existence and their typical characteristics associated with the
different types of trends. In addition, we also show how to use
DFA appropriately to minimize or even eliminate the effects of
those trends in cases that trends are not choices of the study,
that is, trends do not reflect the dynamics of the system but are
caused by some ``irrelevant'' background. Finally,
Sec.~\ref{seccon} contains a summary.

\section{DFA}\label{secdfa}
To illustrate the DFA method, we consider a noisy time series,
$u(i)$ ($i=1,..,N_{max}$~). We integrate the time series $u(i)$,
\begin{equation}
y(j) = \sum\limits_{i=1}^{j} (u(i) - <u>),
\end{equation}
where
\begin{equation}
<u>=\frac{1}{N_{max}} \sum\limits_{j=1}^{N_{max}} u(i),
\end{equation}
and is divided into boxes of equal size, $n$. In each box, we
fit the integrated time series by using a polynomial function,
$y_{fit}(i)$, which is called the local trend. For order-$\ell$ DFA
(DFA-1 if $\ell=1$, DFA-2 if $\ell=2$ etc.), $\ell$ order polynomial function
should be applied for the fitting.  We detrend The integrated time series,
$y(i)$ by subtracting the local trend $y_{fit}(i)$
in each box, and we calculate the detrended fluctuation function
\begin{equation}
Y(i) = y(i)-y_{fit}(i).
\label{psi}
\end{equation}
For a given box size $n$, we calculate the root mean square (rms) fluctuation
\begin{equation}
F(n) =\sqrt{\frac{1}{N_{max}}\sum\limits_{i=1}^{N_{max}}\left [Y(i)\right ]^2}
\label{F}
\end{equation}
The above computation is repeated for box
sizes $n$ (different scales) to provide a relationship between $F(n)$ and $n$. A
power-law relation between $F(n)$ and the box size $n$ indicates
the presence of scaling: $F(n) \sim n^{\alpha}$. The parameter
$\alpha$, called the scaling exponent or correlation exponent, represents the correlation
properties of the signal: if $\alpha=0.5$, there is no correlation
and the signal is an uncorrelated signal (white noise); if
$\alpha < 0.5$, the signal is anticorrelated; if
$\alpha >0.5$, there are positive correlations in the signal.

\section{Noise with linear trends}\label{seclin}
First we consider the simplest case: correlated noise with a
linear trend. A linear trend
\begin{equation}
u(i)=A_{\rm L} i
\end{equation}
is characterized by
only one variable --- the slope of the trend, $A_{\rm L}$. For
convenience, we denote the rms fluctuation function
for noise without trends by $F_{\rm \eta}(n)$, linear trends by
$F_{\rm L}(n)$, and noise with a linear trend by $F_{\rm
\eta L}(n)$.

\subsection{DFA-1 on noise with a linear trend}\label{secdfa1lb}
Using the algorithm of Makse\cite{MFFM}, we generate correlated noise with standard deviation one, with a
given correlation property characterized by a given scaling
exponent $\alpha$. We apply DFA-1 to quantify the correlation
properties of the noise and find that only in certain good fit
region the rms fluctuation function $F_{\rm \eta}(n)$
can be approximated by a power-law function
[see Appendix \ref{secpuren}]
\begin{equation}
F_{\rm \eta}(n) = b_0 n^{\alpha} \label{dfa1_n}
\end{equation}
where $b_0$ is a parameter independent of the scale $n$. We find
that the good fit region depends on the correlation exponent
$\alpha$ 
[see Appendix \ref{secpuren}]. We also
derive analytically the rms fluctuation function for
linear trend only for DFA-1 and find that [see Appendix
\ref{secdfa1l}]
\begin{equation}
F_{\rm L}(n) = k_0 A_{\rm L}  n^{\alpha_{L}}
\label{dfa1_purelb}
\end{equation}
where $k_0$ is a constant independent of the length of trend $N_{max}$, of
the box size $n$ and of the slope of the trend $A_{\rm L}$. We obtain
$\alpha_{L} = 2$.
\begin{figure}[H!]
\centerline{
\epsfysize=0.47\textwidth{\rotate[r]{\epsfbox{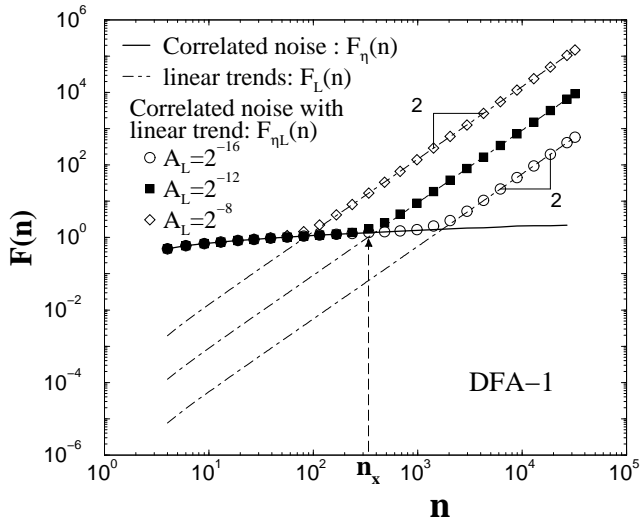}}}}
\vspace*{0.47cm} \caption{ Crossover behavior of the root mean
square fluctuation function $F_{\rm \eta L}(n)$ for noise (of
length $N_{max}=2^{17}$ and correlation exponent $\alpha =0.1$)
with superposed linear trends of slope $A_{\rm L}=2^{-16}, 2^{-12},
2^{-8}$. For comparison, we show $F_{\rm \eta}(n)$ for the noise
(thick solid line) and $F_{\rm L}(n)$ for the linear trends
(dot-dashed line) (Eq.(\ref{dfa1_purelb})). The results show that
a crossover at a scale $n_{\times}$ for $F_{\rm \eta L}(n)$.  For
$n < n_{\times}$,  the noise  dominates and $F_{\rm \eta L}(n)
\approx F_{\rm \eta}(n)$. For $n > n_{\times}$, the linear trend
dominates and $F_{\rm \eta L}(n) \approx F_{\rm L}(n)$. Note that
the crossover scale $n_{\times}$ increases when the slope $A_{\rm L}$
of the trend decreases. } \label{dfa1_npbl_r_a01n17}
\end{figure}
Next we apply the DFA-1 method to the superposition of a linear trend
with correlated noise and we compare the rms fluctuation
function $F_{\rm \eta L}(n)$ with $F_{\rm \eta}(n)$ [see
Fig.\ref{dfa1_npbl_r_a01n17}]. We observe a crossover in $F_{\rm
\eta L}(n)$ at scale $n = n_{\times}$. For
$n<n_{\times}$, the behavior of $F_{\rm \eta L}(n)$ is very close
to the behavior of $F_{\rm \eta}(n)$, while for $n
>n_{\times}$, the behavior of $F_{\rm \eta L}(n)$ is very close to the behavior of $F_{\rm L}(n)$.
A similar crossover behavior is also observed in the scaling of the
well-studied biased random walk \cite{Weissrdwalk,newtontoman}. It
is known that the crossover in the biased random walk is due to the
competition of the unbiased random walk and the bias [see Fig.5.3 of \cite{newtontoman}]. We
illustrate this observation in Fig.~\ref{x_dfa1_addlb_32_a01n17},
where the detrended fluctuation functions (Eq.~(\ref{psi})) of the
correlated noise, $Y_{\rm \eta}(i)$, and of the noise with a
linear trend, $Y_{\rm \eta L}(i)$ are shown. For the box size $
n < n_{\times}$ as shown in Fig.~\ref{x_dfa1_addlb_32_a01n17}(a)
and (b), $Y_{\rm \eta L}(i) \approx Y_{\rm \eta}(i)$. For
$n>n_{\times}$ as shown in Fig.~\ref{x_dfa1_addlb_32_a01n17}(c)
and (d), $Y_{\rm \eta L}(i)$ has distinguishable quadratic
background significantly different from $Y_{\rm \eta}(i)$. This
quadratic background is due to the integration of the linear trend
within the DFA procedure and represents the detrended fluctuation
function $Y_{L}$ of the linear trend. These relations between
the detrended fluctuation functions $Y(i)$ at different time
scales $n$ explain the crossover in the scaling behavior of
$F_{\rm \eta L}(n)$: from very close to $F_{\rm \eta}(n)$ to very close
to $F_{\rm L}(n)$ (observed in Fig.\ref{dfa1_npbl_r_a01n17}).

\begin{figure}[H!]
\centerline{
\epsfysize=0.47\textwidth{\rotate[r]{\epsfbox{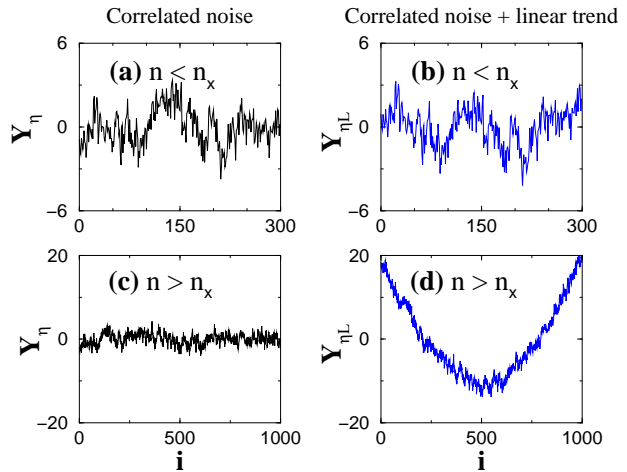}}}}
\vspace*{0.25cm} \caption{ Comparison of the detrended fluctuation
function for noise $Y_{\rm \eta}(i)$ and for noise with linear
trend $Y_{\rm \eta L}(i)$ at different scales. (a) and (c) are
$Y_{\rm \eta}$ for noise with $\alpha = 0.1$; (b) and (d) are
$Y_{\rm \eta L}$ for the same noise with a linear trend with
slope $A_{\rm L}=2^{-12}$ (the crossover scale $n_{\times} = 320$ see
Fig.~\ref{dfa1_npbl_r_a01n17}). (a) (b) for scales $n <
n_{\times}$ the effect of the trend is not pronounced and
$Y_{\rm \eta} \approx Y_{\rm \eta L}$ (i.e. $Y_{\rm
\eta} \gg Y_{\rm L}$); (c)(d) for scales $n > n_{\times}$, the
linear trend is dominant and $Y_{\rm \eta} \ll Y_{\rm \eta L}$. } \label{x_dfa1_addlb_32_a01n17}
\end{figure}

The experimental results presented in Figs.\ref{dfa1_npbl_r_a01n17} and \ref{x_dfa1_addlb_32_a01n17} suggest that the rms fluctuation function for a signal which is a superposition of a correlated noise and a linear trend can be expressed as:
\begin{equation}
\left [F_{\rm \eta L}(n)\right ]^2 = \left [F_{\rm L}(n)\right ]^2 + \left [F_{\rm \eta}(n)\right ]^2
\label{addnl}
\end{equation}
We provide an analytic derivation of this relation in Appendix \ref{secadd}, where we show that Eq.(\ref{addnl}) holds for the superposition of any two independent signals --- in this particular case noise and a linear trend. We call this relation the ``superposition rule''. This rule helps us understand how the competition between the contribution of the noise and the trend to the rms fluctuation function $F_{\rm \eta L}(n)$ at different scales $n$ leads to appearance of crossovers \cite{Weissrdwalk}.

\begin{figure}[H!]
\centerline{
\epsfysize=0.47\textwidth{\rotate[r]{\epsfbox{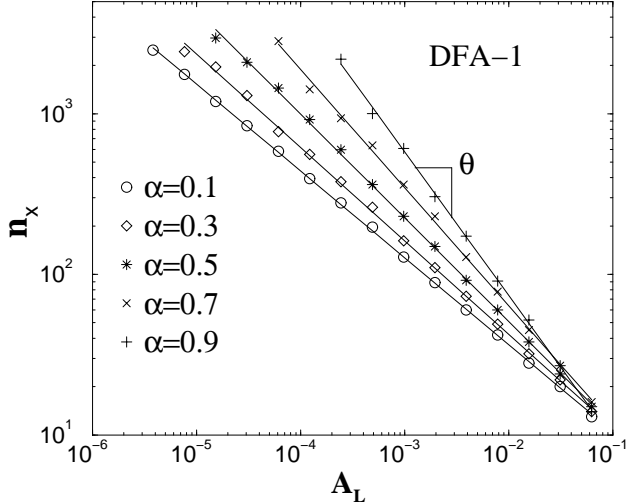}}}}
\vspace*{0.5cm} \caption{The crossover $n_{\times}$ of $F_{\rm
\eta L}(n)$ for noise with a linear trend. We determine the
crossover scale $n_{\times}$ based on the difference $\Delta$
between $\log F_{\rm \eta}$ (noise) and $\log F_{\rm \eta L}$
(noise with a linear trend). The scale for which $\Delta =0.05$ is
the estimated crossover scale $n_{\times}$. For any given
correlation exponent $\alpha$ of the noise, the crossover scale
$n_{\times}$ exhibits a long-range power-law behavior $n_{\times}
\sim \left(A_{\rm L}\right)^{\theta}$, where the crossover exponent $\theta$ is
a function of $\alpha$ [see Eq.(\ref{si_bl}) and Table
\ref{slopefit}].} \label{S_r_dfa1_nbl_n17}
\end{figure}

Next, we ask how the crossover scale $n_{\times}$ depends on: (i)
the slope of the linear trend $A_{\rm L}$, (ii) the scaling exponent
$\alpha$ of the noise, and (iii) the length of the signal
$N_{max}$. Surprisingly, we find that for noise with any given
correlation exponent $\alpha$ the crossover scale $n_{\times}$
itself follows a power-law scaling relation over several decades:
$ n_{\times} \sim \left(A_{\rm L}\right)^{\theta}$ (see
Fig.~\ref{S_r_dfa1_nbl_n17}). We find that in this scaling
relation, the crossover exponent $\theta$ is negative and its value
depends on the correlation exponent $\alpha$ of the noise --- the
magnitude of $\theta$ decreases when $\alpha$ increases. We present the
values of the ``crossover exponent'' $\theta$ for different correlation
exponents $\alpha$ in Table~\ref{slopefit}.

\begin{minipage}[t]{0.9\columnwidth}
\begin{table}[H!]
\caption{ The crossover exponent $\theta$ from the power-law relation between the crossover scale $n_{\times}$ and the slope of the linear trend $A_{\rm L}$ --- $n_{\times} \sim \left(A_{\rm L}\right)^{\theta}$ ---for different values of the correlation exponents $\alpha$ of the noise [Fig.~\ref{S_r_dfa1_nbl_n17}]. The values of $\theta$ obtained from our simulations are in good agreement with the analytical prediction $-1/(2-\alpha)$ [Eq.~(\ref{si_bl})]. Note that $-1/(2-\alpha)$ are not always exactly equal to $\theta$ because $F_{\rm \eta}(n)$ in simulations is not a perfect simple power-law function and the way we determine numerically $n_{\times}$ is just approximated.}\label{slopefit}
\begin{tabular}{@{}c@{~~~~~~~~~}c@{~~~~~~}c@{}}
\centering
$\alpha   $ & $\theta$ & $-1/(2-\alpha)$ \\ \tableline
  0.1 & -0.54 & -0.53 \\
  0.3 & -0.58 & -0.59 \\
  0.5 & -0.65 & -0.67 \\
  0.7 & -0.74 & -0.77 \\
  0.9 & -0.89 & -0.91 \\
\end{tabular}
\end{table}
\end{minipage}

To understand how the crossover scale depends on the correlation exponent $\alpha$ of the noise we employ the superposition rule [Eq.(\ref{addnl})] and estimate $n_{\times}$ as the intercept between $F_{\rm \eta}(n)$ and $F_{\rm L}(n)$. From the Eqs.~(\ref{dfa1_n}) and (\ref{dfa1_purelb}), we obtain the following dependence of $n_{\times}$ on $\alpha$:
\begin{equation}
n_{\times} = \left(A_{\rm L}\frac{k_0}{b_0}\right)^{1/(\alpha-\alpha_{L})} = \left(A_{\rm L}\frac{k_0}{b_0}\right)^{1/(\alpha-2)}
\label{si_bl}
\end{equation}
This analytical calculation for the crossover exponent $-1/(\alpha_{L}-\alpha)$ is in a good agreement with the observed values of $\theta$ obtained from our simulations [see Fig.\ref{S_r_dfa1_nbl_n17} and Table \ref{slopefit}].

Finally, since the $F_{\rm L}(n)$ does not depend on $N_{max}$ as
we show in Eq.(\ref{dfa1_purelb}) and in Appendix \ref{secdfa1l},
we find that $n_{\times}$ does not depend on $N_{max}$. This is a
special case for linear trends and does not always hold for higher
order polynomial trends [see Appendix \ref{secq}].

\subsection{DFA-2 on noise with a linear trend}\label{secdfa2bl}
\begin{figure}[H!]
\centerline{
\epsfysize=0.47\textwidth{\rotate[r]{\epsfbox{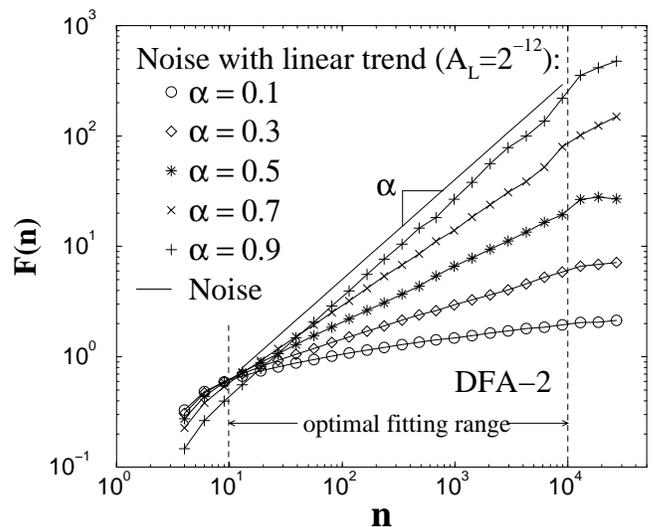}}}}
\vspace*{0.5cm} \caption{Comparison of the rms fluctuation
function $F_{\rm \eta}(n)$ for noise with different types of
correlations (lines) and $F_{\rm \eta L}(n)$ for the same noise
with a linear trend of slope $A_{\rm L} =2^{-12}$ (symbols) for DFA-2.
$F_{\rm \eta L}(n)=F_{\rm \eta}(n)$ because the integrated linear
trend can be perfectly filtered out in DFA-2, thus $Y_{\rm L}(i)
= 0$ from Eq.(\ref{psi}). We note, that to estimate accurately the
correlation exponents one has to choose an optimal range of scales
$n$, where $F(n)$ is fitted. For details see
Appendix~\ref{secpuren}}. \label{dfa2_nbl_32_n17FT}
\end{figure}
Application of the DFA-2 method to noisy signals without any
polynomial trends leads to scaling results identical to the
scaling obtained from the DFA-1 method, with the exception of some
vertical shift to lower values for the rms fluctuation function
$F_{\rm \eta}(n)$ [see Appendix~\ref{secpuren}]. However, for
signals which are a superposition of correlated noise and a linear
trend, in contrast to the DFA-1 results presented in
Fig.~\ref{dfa1_npbl_r_a01n17}, $F_{\rm \eta L}(n)$ obtained from
DFA exhibits no crossovers, and is exactly equal to the rms
fluctuation function $F_{\rm \eta}(n)$ obtained from DFA-2 for
correlated noise without trend (see Fig.~\ref{dfa2_nbl_32_n17FT}).
These results indicate that a linear trend has no effect on the
scaling obtained from DFA-2. The reason for this is that by design
the DFA-2 method filters out linear trends, i.e. $Y_{\rm L}(i) =
0$ (Eq.(~\ref{psi})) and thus $F_{\rm \eta L}(n)=F_{\rm \eta}(n)$
due to the superposition rule (Eq.~(\ref{addnl})). For the same
reason, polynomial trends of order lower than $\ell$ superimposed on
correlated noise will have no effect on the scaling properties of
the noise when DFA-$\ell$ is applied. Therefore, our results
confirm that the DFA method is a reliable tool to accurately
quantify correlations in noisy signals embedded in polynomial
trends. Moreover, the reported scaling and crossover features of
$F(n)$ can be used to determine the order of polynomial trends
present in the data.

\section{Noise with sinusoidal trend } \label{secsin}

In this section, we study the effect of sinusoidal trends on the
scaling properties of noisy signals. For a signal which is a
superposition of correlated noise and sinusoidal trend, we find
that based on the superposition rule (Appendix~\ref{secadd}) the
DFA rms fluctuation function can be expressed as
\begin{equation}
\left[F_{\rm \eta S}(n)\right]^2 = \left[F_{\rm \eta}(n)\right]^2 + \left[F_{\rm S}(n)\right]^2,
\label{addnp}
\end{equation}
where $F_{\rm \eta S}(n)$ is the rms fluctuation
function of noise with a sinusoidal trend, and $F_{\rm S}(n)$ is
for the sinusoidal trend. First we consider the application of DFA-1 to a sinusoidal trend. Next we study the scaling behavior and the features of crossovers in $F_{\rm
\eta \rm S}(n)$ for the superposition of correlated noise and sinusoidal trend
employing the superposition rule [Eq.(\ref{addnp})]. At the end
of this section, we discuss the results obtained from higher order DFA.

\subsection{DFA-1 on sinusoidal trend}\label{secsin_puresin}

\begin{figure}[H!]
\centerline{
\epsfysize=0.47\textwidth{\rotate[r]{\epsfbox{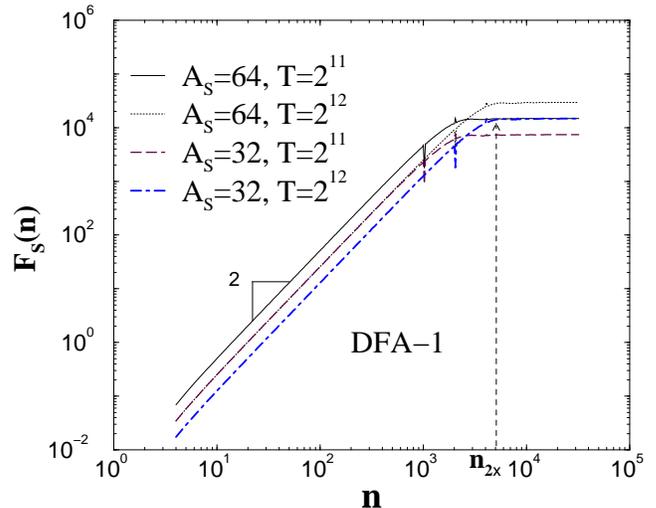}}}}
\caption{Root mean square fluctuation function $F_{\rm S}(n)$ for
sinusoidal functions of length $N_{max}=2^{17}$ with different
amplitude $A_{\rm S}$ and period $T$. All curves exhibit a
crossover at $n_{2\times} \approx T/2$, with a slope $\alpha_{\rm
S}=2$ for $n < n_{2\times}$, and a flat region for $n
> n_{2\times}$. There are some spurious singularities at $n = j
\frac{T}{2}$ ($j$ is a positive integer) shown by the spikes.}
\label{dfa1_puresin}
\end{figure}

Given a sinusoidal trend $u(i)= A_{\rm S} \sin \left(2\pi i/T\right)$
($i=1,...,N_{max}$), where $A_{\rm S}$ is the amplitude of the
signal and $T$ is the period, we find that the rms fluctuation
function $F_{\rm S}(n)$ does not depend on the length of the
signal $N_{max}$, and has the same shape for different amplitudes
and different periods [Fig.~\ref{dfa1_puresin}]. We find a
crossover at scale corresponding to the period of the sinusoidal
trend
\begin{equation}
n_{2\times} \approx T, \label{eqnx2}
\end{equation}
and does not depend on the amplitude $A_{\rm S}$. We
call this crossover  $n_{2 \times}$ for convenience, as we will see later. For $n~<~
n_{2\times}$, the rms fluctuation $F_{\rm S}(n)$ exhibits an
apparent scaling with the same exponent as $F_{\rm L}(n)$ for the
linear trend [see Eq.~(\ref{dfa1_purelb})]:
\begin{equation}
F_{\rm S}(n) = k_1 \frac{A_{\rm S}}{T} n^{\alpha_{\rm S}}
\label{eqpl}
\end{equation}
where $k_1$ is a constant independent of the length $N_{max}$, of
the period $T$ and the amplitude $A_{\rm S}$ of the sinusoidal
signal, and of the box size $n$. As for the linear trend
[Eq.(\ref{dfa1_purelb})], we obtain $\alpha_{\rm S}~=~2$ because
at small scales (box size $n$) the sinusoidal function is
dominated by a linear term. For $n > n_{2\times}$, due to the
periodic property of the sinusoidal trend, $F_{\rm S}(n)$ is a
constant independent of the scale $n$:
\begin{equation}
F_{\rm S}(n) = \frac{1}{2\sqrt{2} \pi} A_{\rm S} \cdot T.
\label{eqpp}
\end{equation}
The period $T$ and the amplitude $A_{\rm S}$ also affects the
vertical shift of $F_{\rm S}(n)$ in both regions.  We note that in
Eqs.(\ref{eqpl}) and (\ref{eqpp}), $F_{\rm S}(n)$ is proportional to the
amplitude $A_{\rm S}$, a behavior which is also observed for the
linear trend [Eq.~(\ref{dfa1_purelb})].

\subsection{DFA-1 on noise with sinusoidal trend}\label{secdfa1st}

\begin{figure}[H!]
\centerline{
\epsfysize=0.43\textwidth{\rotate[r]{\epsfbox{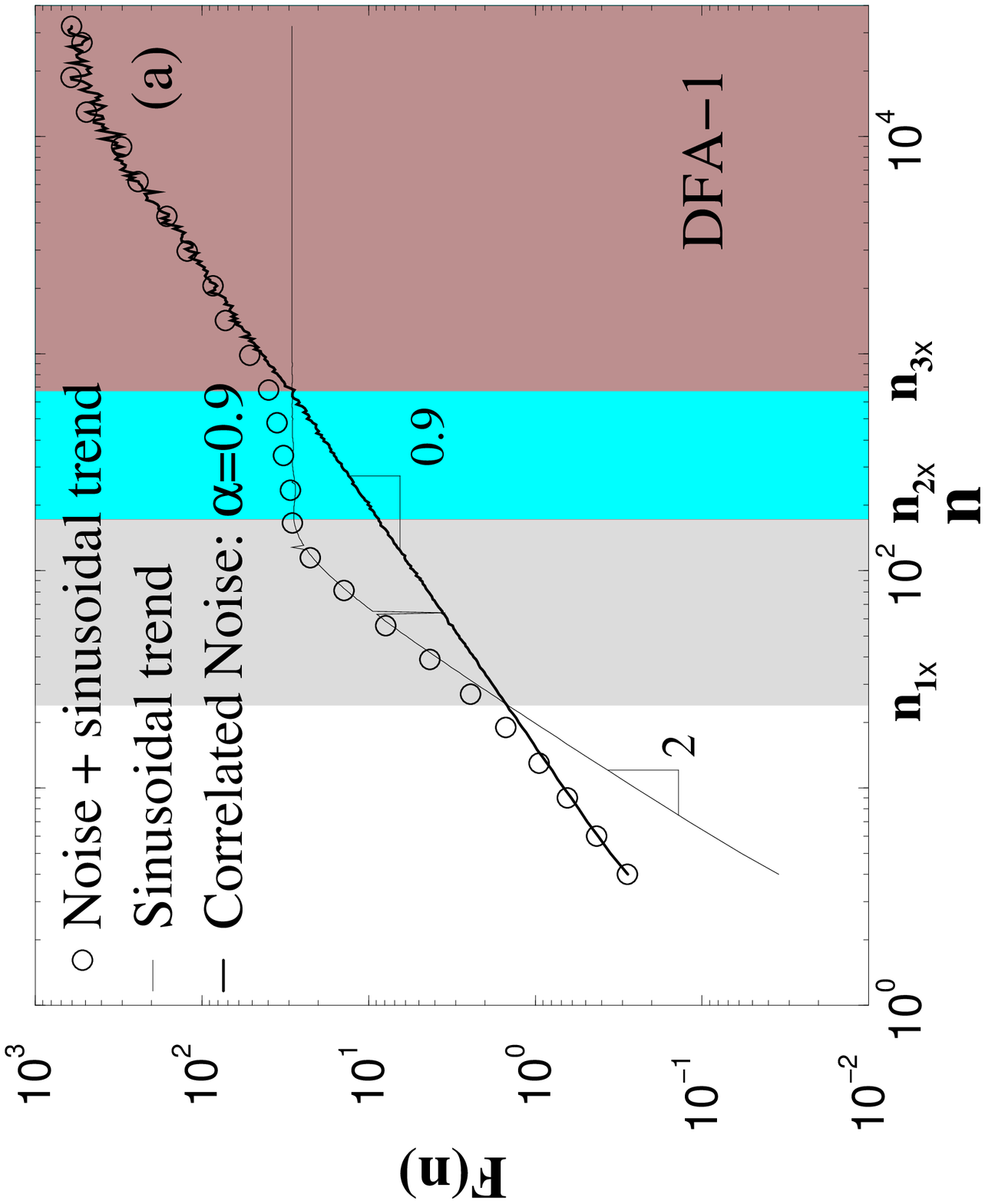}}}
} \centerline{
\epsfysize=0.43\textwidth{\rotate[r]{\epsfbox{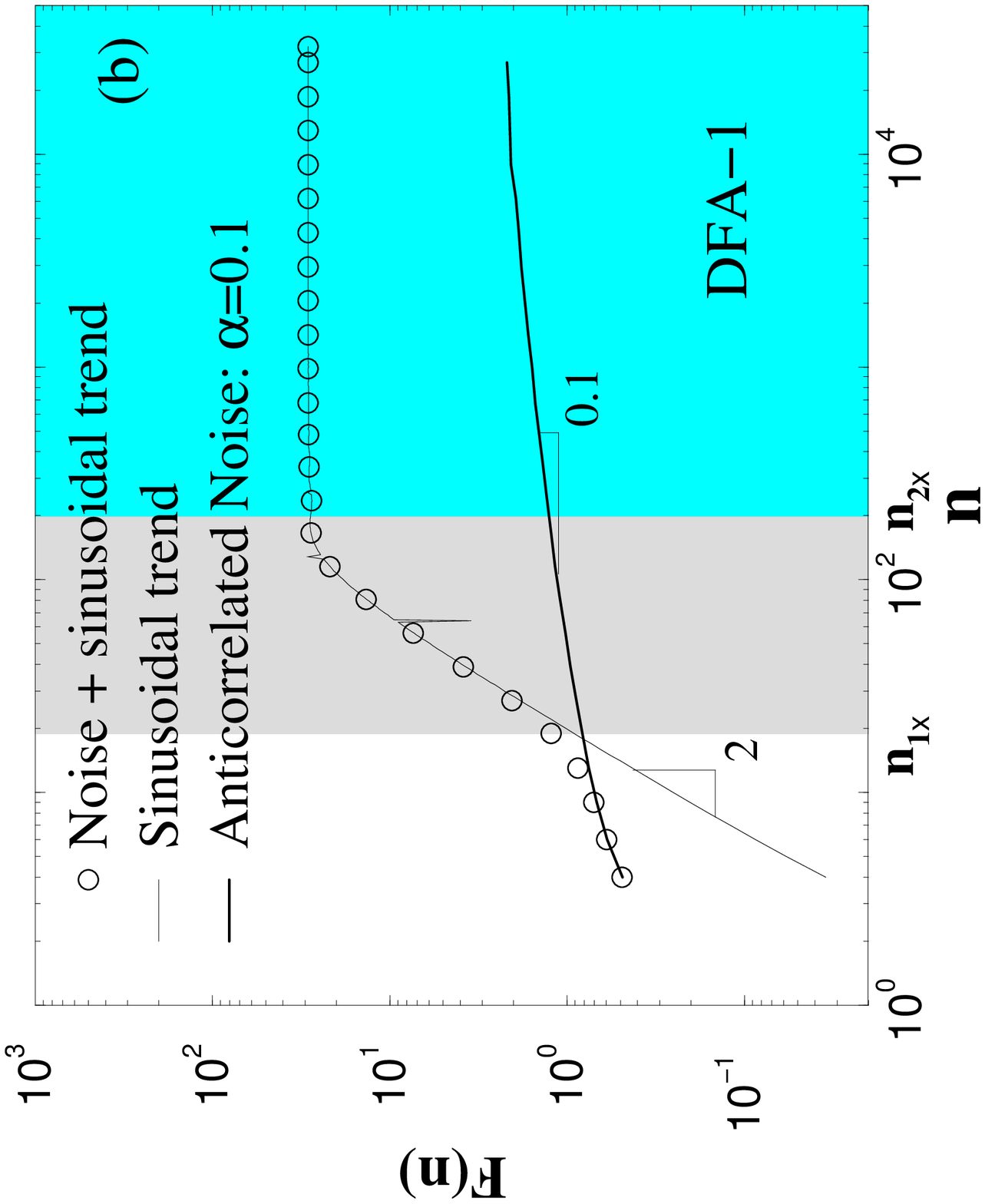}}}}
\vspace*{0.25cm} \caption{Crossover behavior of the root mean
square fluctuation function $F_{\rm \eta \rm S}(n)$ (circles) for
correlated noise (of length $N_{max}=2^{17}$) with a superposed
sinusoidal function characterized by period $T=128$ and amplitude
$A_{\rm S} = 2$. The rms fluctuation function $F_{\rm
\eta}(n)$ for noise (thick line) and $F_{\rm S}(n)$ for the
sinusoidal trend (thin line) are shown for comparison. (a) $F_{\rm
\eta \rm S}(n)$ for correlated noise with $\alpha=0.9$. (b)
$F_{\rm \eta \rm S}(n)$ for anticorrelated noise with
$\alpha=0.9$. There are three crossovers in $F_{\rm \eta \rm
S}(n)$, at scales $n_{1\times}$, $n_{2\times}$ and $n_{3\times}$
(the third crossover can not be seen in (b) because it occurs at
scale larger than the length of the signal). For $n < n_{1\times}$
and $n > n_{3\times}$, the noise dominates and $F_{\rm \eta \rm
S}(n) \approx F_{\rm \eta}(n)$ while for $ n_{1\times} < n <
n_{3\times}$, the sinusoidal trend dominates and $F_{\rm \eta \rm
S}(n) \approx F_{\rm S}(n)$. The crossovers at $n_{1\times}$ and
$n_{3\times}$ are due to the competition between the correlated
noise and the sinusoidal trend [see Fig.~\ref{x_nbw_2_128_n17}],
while the crossover at $n_{2\times}$ relates only to the period
$T$ of the sinusoidal [Eq.~(\ref{eqnx2})].} \label{dfa1_nbw_n17}
\end{figure}

\begin{figure}[ht]
\centerline{
\epsfysize=0.47\textwidth{\rotate[r]{\epsfbox{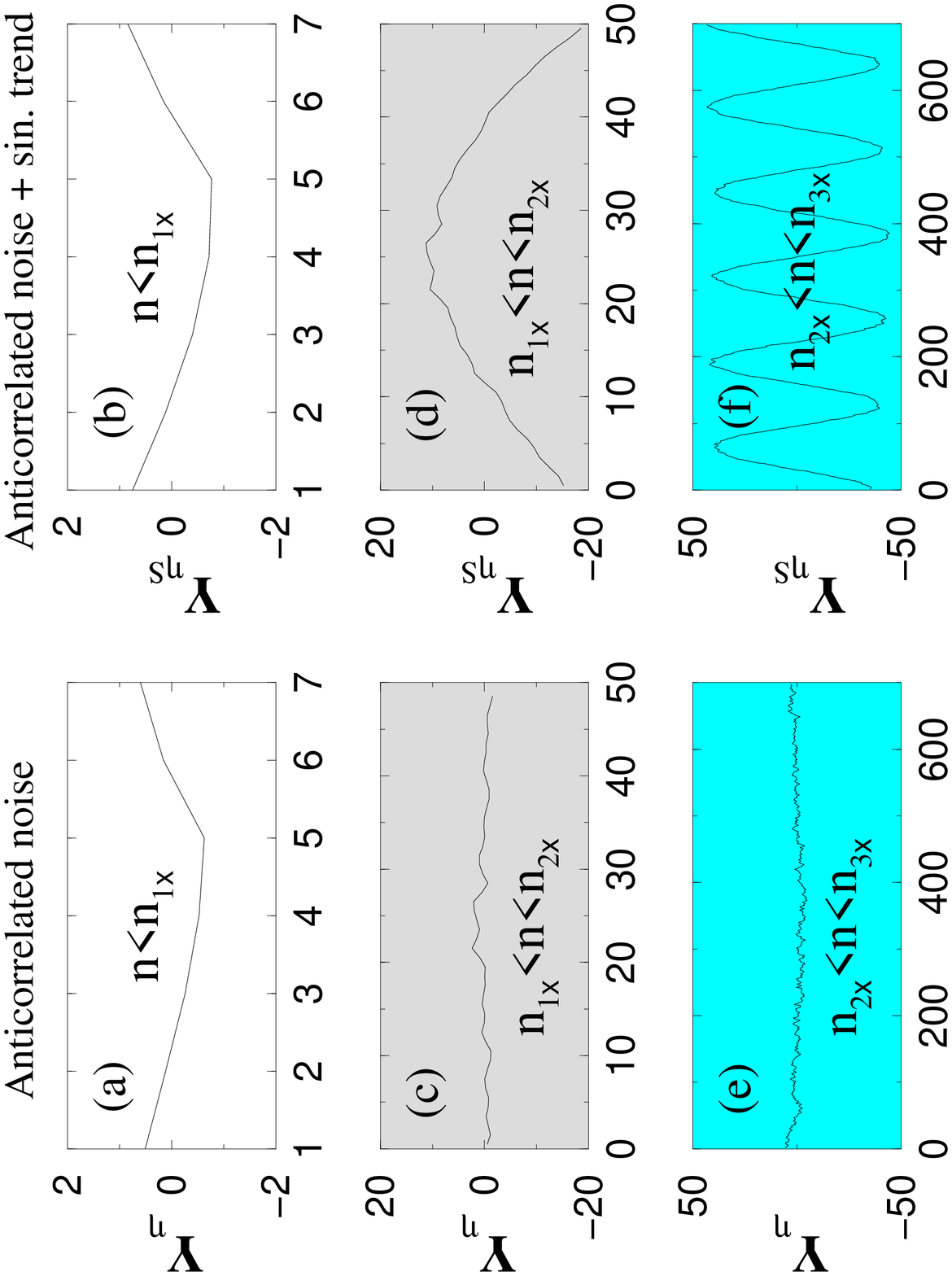}}}}
\centerline{
\epsfysize=0.47\textwidth{\rotate[r]{\epsfbox{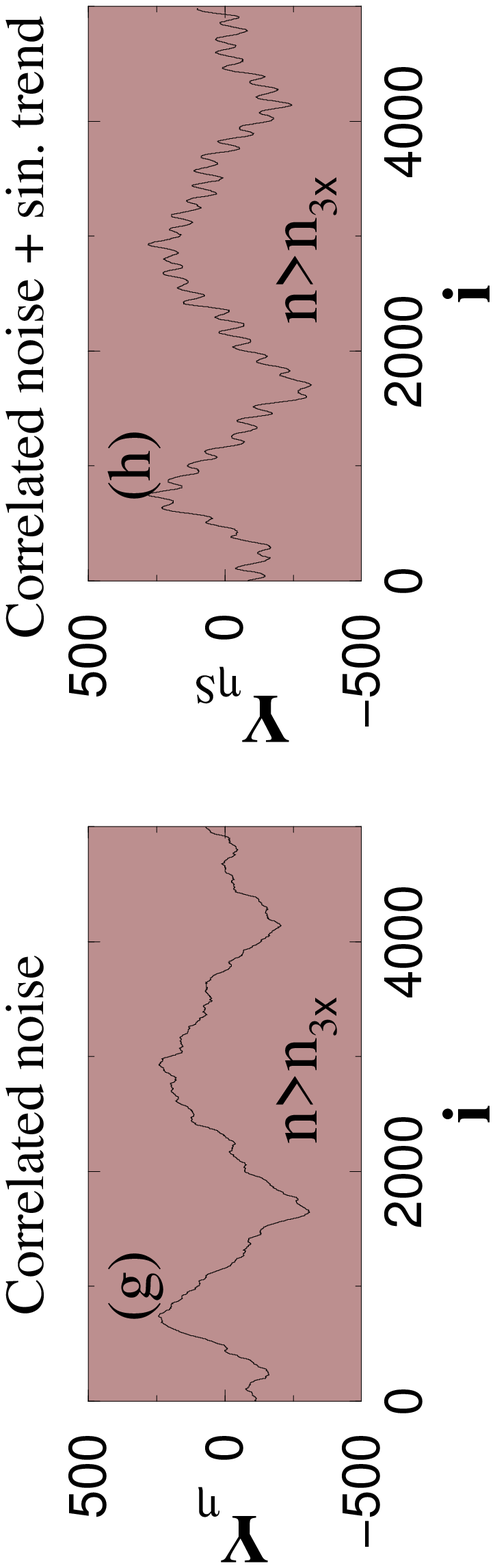}}}}
\vspace*{0.25cm} \caption{Comparison of the detrended fluctuation
function for noise, $Y_{\rm \eta}(i)$ and noise with sinusoidal
trend, $Y_{\rm \eta S}(i)$ in four regions as shown in
Fig.~\ref{dfa1_nbw_n17}. The same signals as in
Fig.~\ref{dfa1_nbw_n17} are used. Panels (a)-(f) correspond to
Fig.~\ref{dfa1_nbw_n17}(b) for anticorrelated noise with exponent
$\alpha=0.1$, and panels (g)-(h) correspond to the
Fig.~\ref{dfa1_nbw_n17}(a) for correlated noise with exponent
$\alpha=0.9$. (a)-(b) For all scales $ n < n_{1\times}$, the effect of
the trend is not pronounced and $Y_{\rm \eta \rm S}(i) \approx
Y_{\rm \eta}(i)$ leading to $F_{\rm \eta S}(n) \approx F_{\rm
\eta}(n)$ (Fig.~\ref{dfa1_nbw_n17}(a)). (c)(d) For $ n_{2\times}>
n
> n_{1\times}$, the trend is dominant, $Y_{\rm \eta S}(i)
\gg Y_{\rm \eta}(i)$ and $F_{\rm \eta S}(n)\approx F_{\rm
S}(n)$. Since $n_{2\times} \approx T/2$ (Eq.~(\ref{eqnx2})), the
scale $n < T/2$ and the sinusoidal behavior can be approximated as
a linear trend. This explains the quadratic background in
$Y_{\rm \eta S}(i)$ (d) [see
Fig.~\ref{x_dfa1_addlb_32_a01n17}(c)(d)]. (e)(f) For $n_{2\times}
< n < n_{3\times}$ (i.e. $n\gg T/2$), the sinusoidal trend again
dominates --- $Y_{\rm \eta S}(i)$ is periodic function with
period $T$. (g)(h) for $n~>~n_{3\times}$, the effect of the noise
is dominant and the scaling of $F_{\rm \eta S}$ follows the
scaling of $F_{\rm \eta}$ (Fig.~\ref{dfa1_nbw_n17}(a)). }
\label{x_nbw_2_128_n17}
\end{figure}

In this section, we study how the sinusoidal trend affects the
scaling behavior of noise with different type of correlations. We
apply the DFA-1 method to a signal which is a superposition of
correlated noise with a sinusoidal trend. We observe that there
are typically three crossovers in the rms fluctuation $F_{\rm
\eta S}(n)$ at characteristic scales denoted by $n_{1\times}$,
$n_{2\times}$ and $n_{3\times}$ [Fig.~\ref{dfa1_nbw_n17}]. These
three crossovers divide $F_{\rm \eta S}(n)$ into four regions, as
shown in Fig.~\ref{dfa1_nbw_n17}(a) (the third crossover cannot be
seen in Fig.~\ref{dfa1_nbw_n17}(b) because its scale $n_{3\times}$
is greater than the length of the signal). We find that the first
and third crossovers at scales $n_{1\times}$ and $n_{3\times}$
respectively [see Fig.~\ref{dfa1_nbw_n17}] result from the
competition between the effects on $F_{\rm \eta S}(n)$ of the
sinusoidal signal and the correlated noise. For $n<n_{1\times}$
(region I) and $n>n_{3\times}$ (region IV), we find
that the noise has the dominating effect ($F_{\rm \eta}(n)>F_{\rm
S}(n)$), so the behavior of $F_{\rm \eta S}(n)$ is very close to
the behavior of $F_{\rm \eta}(n)$ [Eq.~(\ref{addnp})]. For
$n_{1\times} < n < n_{2\times}$ (region II) and $n_{2\times} <
n < n_{3\times}$ (region III) the sinusoidal trend dominates
($F_{\rm S}(n)>F_{\rm \eta}(n)$), thus the behavior of $F_{\rm
\eta S}(n)$ is close to $F_{\rm S}(n)$ [see
Fig.~\ref{dfa1_nbw_n17} and Fig.~\ref{x_nbw_2_128_n17}].

To better understand why there are different regions in the
behavior of $F_{\rm \eta \rm S}(n)$, we consider the detrended
fluctuation function [Eq.~(\ref{psi}) and Appendix \ref{secadd}]
of the correlated noise $Y_{\rm \eta}(i)$, and of the noise
with sinusoidal trend $Y_{\rm \eta S}$. In
Fig.~\ref{x_nbw_2_128_n17} we compare $Y_{\rm \eta}(i)$ and
$Y_{\rm \eta S}(i)$ for anticorrelated and correlated noise in
the four different regions. For very small scales $n<n_{1\times}$,
the effect of the sinusoidal trend is not pronounced, $Y_{\rm
\eta S}(i) \approx Y_{\rm \eta}(i)$, indicating that in this
scale region the signal can be considered as noise fluctuating
around a constant trend which is filtered out by the DFA-1
procedure [Fig.~\ref{x_nbw_2_128_n17}(a)(b)]. Note, that the
behavior of $Y_{\rm \eta S}$ [Fig.~\ref{x_nbw_2_128_n17}(b)] is
identical to the behavior of $Y_{\rm \eta L}$
[Fig.~\ref{x_dfa1_addlb_32_a01n17}(b)], since both a sinusoidal
with a large period $T$ and a linear trend with small slope
$A_{\rm L}$ can be well approximated by a constant trend for
$n<n_{1\times}$. For small scales $n_{1\times} < n < n_{2\times}$
(region II), we find that there is a dominant quadratic
background for $Y_{\rm \eta S}(i)$
[Fig.~\ref{x_nbw_2_128_n17}(d)]. This quadratic background is due
to the integration procedure in DFA-1, and is represented by the
detrended fluctuation function of the sinusoidal trend $Y_{\rm
S}(i)$. It is similar to the quadratic background observed for
linear trend $Y_{\rm \eta L}(i)$
[Fig.~\ref{x_dfa1_addlb_32_a01n17}(d)] --- i.e. for $n_{1\times} <
n < n_{2\times}$ the sinusoidal trend behaves as a linear trend
and $Y_{\rm S}(i) \approx Y_{\rm L}(i)$. Thus in region II the ``linear trend'' effect of the sinusoidal is dominant,
$Y_{\rm S}> Y_{\rm \eta}$, which leads to $F_{\rm \eta S}(n)
\approx F_{\rm S}(n)$. This explains also why $F_{\rm \eta S}(n)$
for $n<n_{2\times}$ (Fig.~\ref{dfa1_nbw_n17}) exhibits crossover
behavior similar to the one of $F_{\rm \eta L}(n)$ observed for
noise with a linear trend. For $n_{2\times}<n<n_{3\times}$ (region III) the sinusoidal behavior is strongly pronounced
[Fig.~\ref{x_nbw_2_128_n17}(f)], $Y_{\rm S}(i) \gg Y_{\rm
\eta}(i)$, and $Y_{\rm \eta S}(i) \approx Y_{\rm S}(i)$
changes periodically with period equal to the period of the
sinusoidal trend $T$. Since $Y_{\rm \eta S}(i)$ is bounded
between a minimum and a maximum value, $F_{\rm \eta S}(n)$ cannot
increase 
and exhibits a flat region (Fig.~\ref{dfa1_nbw_n17}). At very
large scales, $n>n_{3\times}$, the noise effect is again dominant
($Y_{\rm S}(i)$ remains bounded, while $Y_{\rm \eta}$ grows
when increasing the scale) which leads to $F_{\rm \eta S}(n)
\approx F_{\rm \eta}(n)$, and a scaling behavior corresponding to
the scaling of the correlated noise.

\begin{figure}[H]
\centerline{
\epsfysize=0.42\textwidth{\rotate[r]{\epsfbox{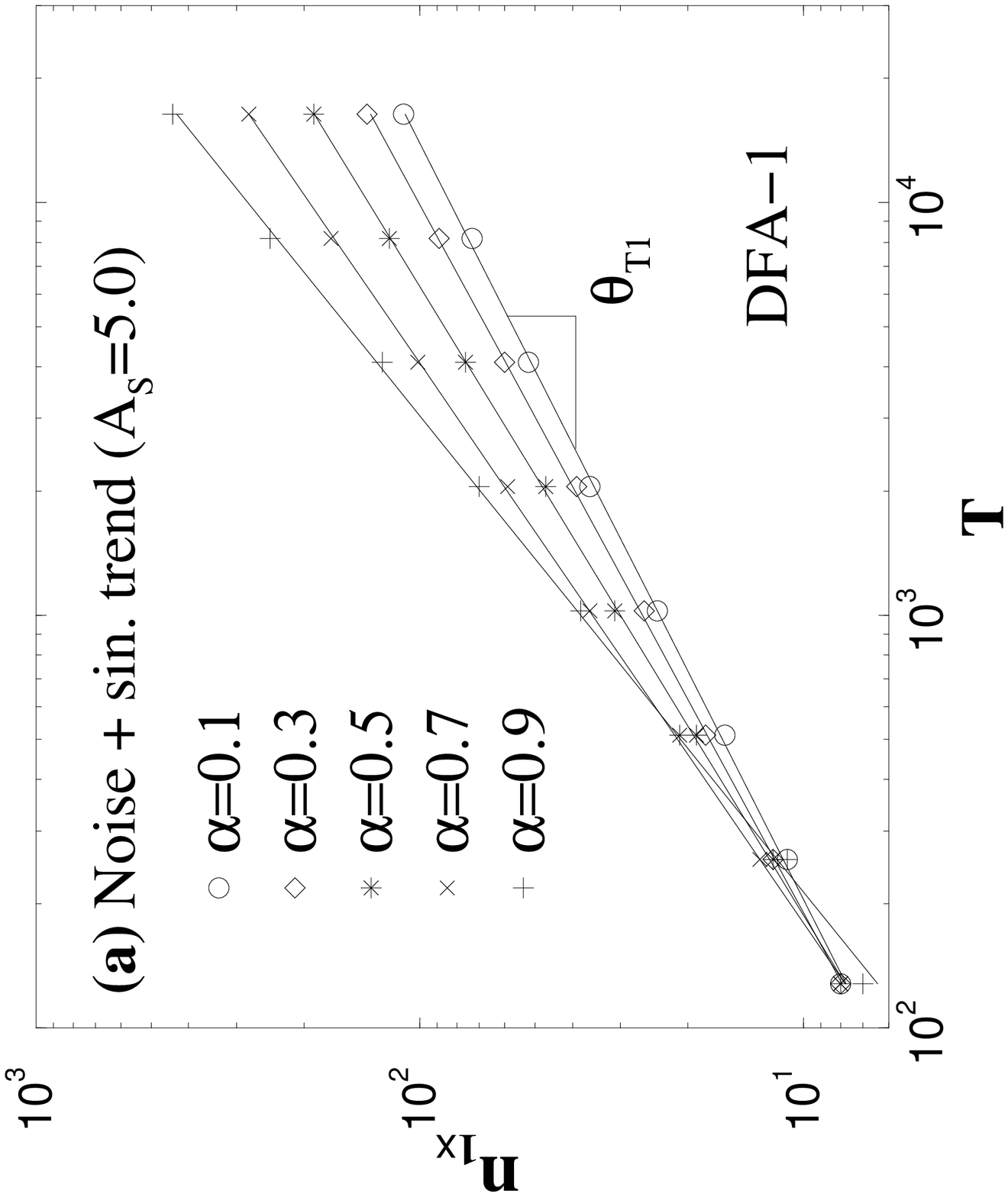}}}
}
\centerline{
\epsfysize=0.42\textwidth{\rotate[r]{\epsfbox{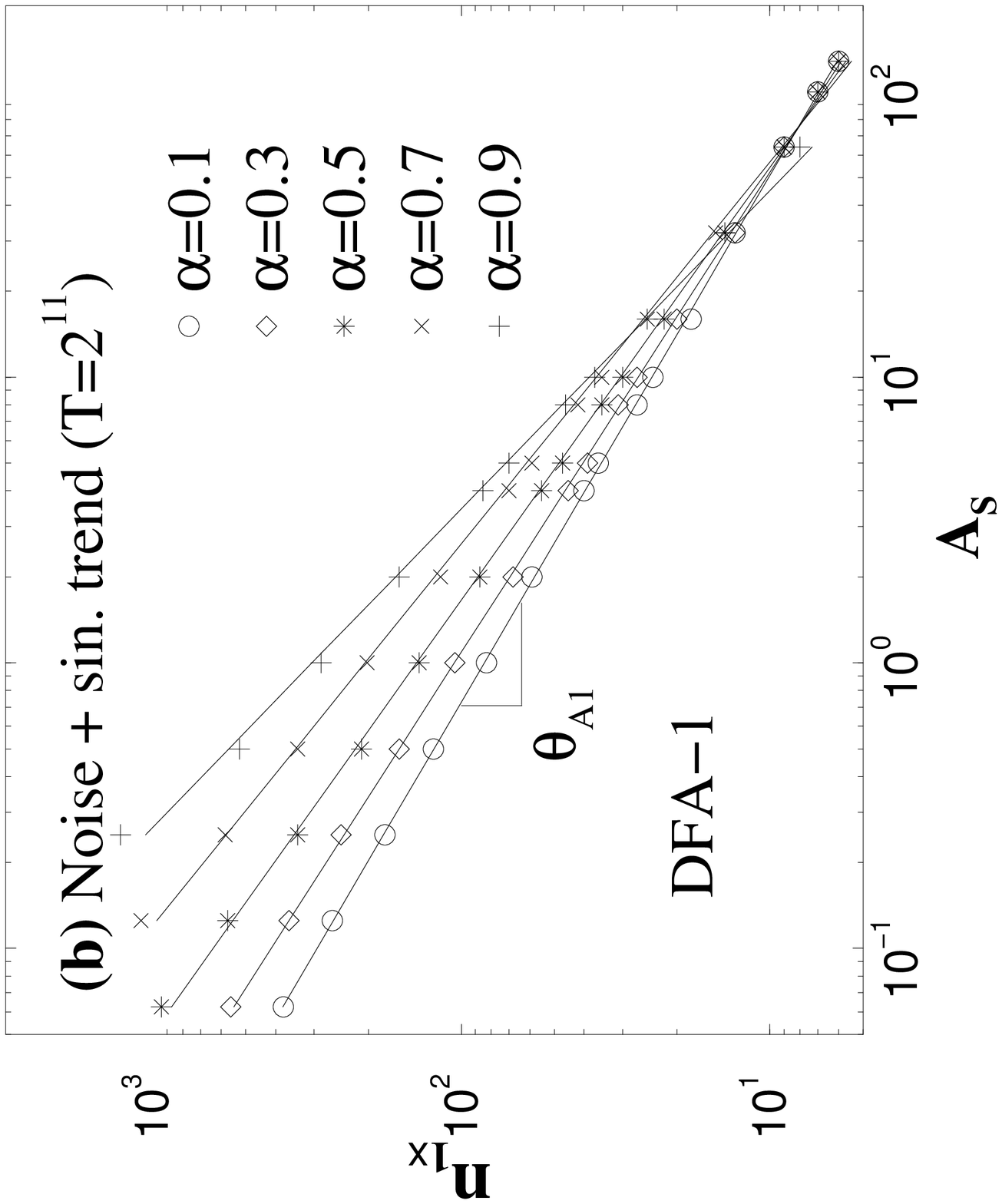}}}}

\centerline{
\epsfysize=0.42\textwidth{\rotate[r]{\epsfbox{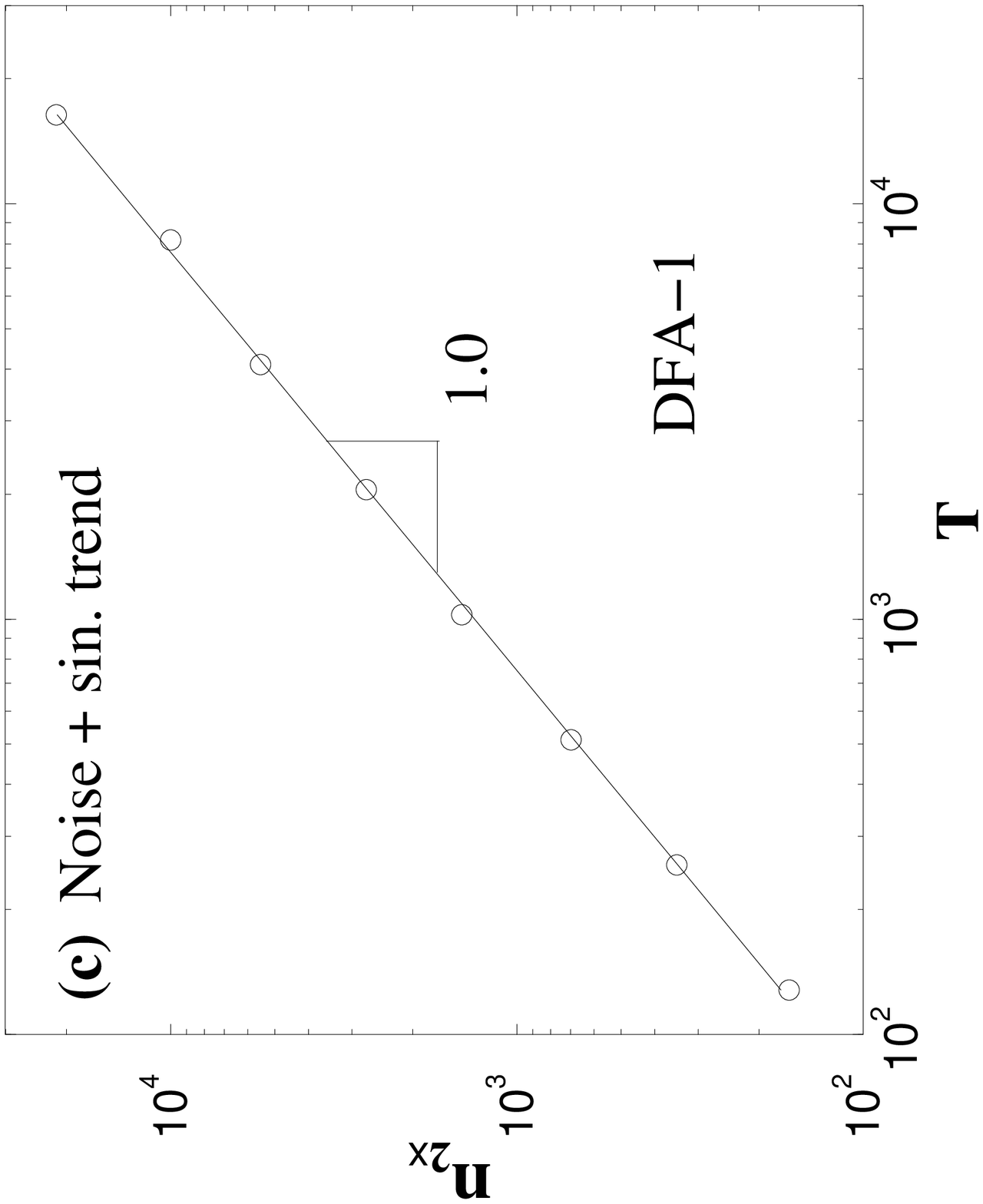}}}}

\centerline{
\epsfysize=0.42\textwidth{\rotate[r]{\epsfbox{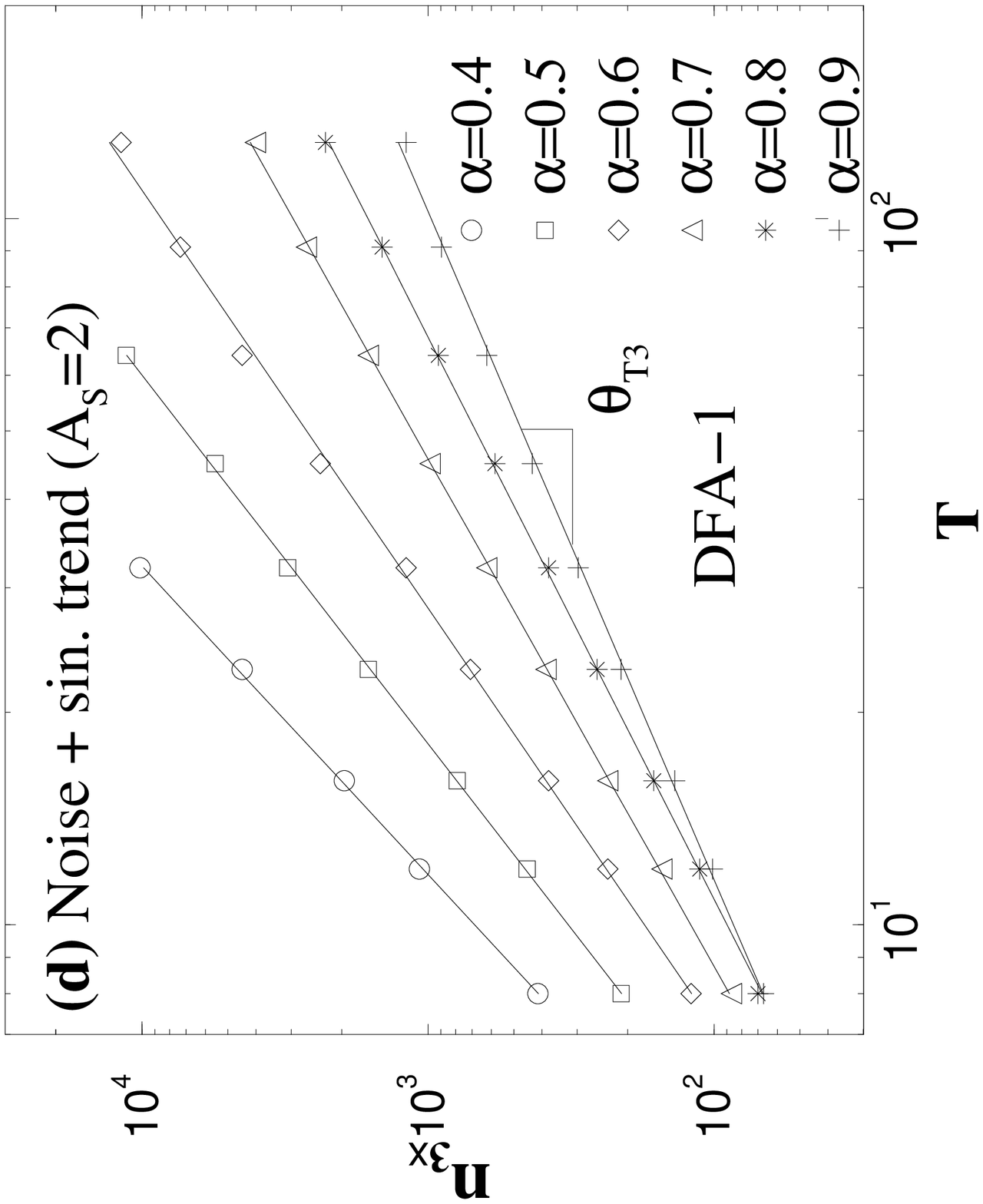}}}
}

\centerline{
\epsfysize=0.42\textwidth{\rotate[r]{\epsfbox{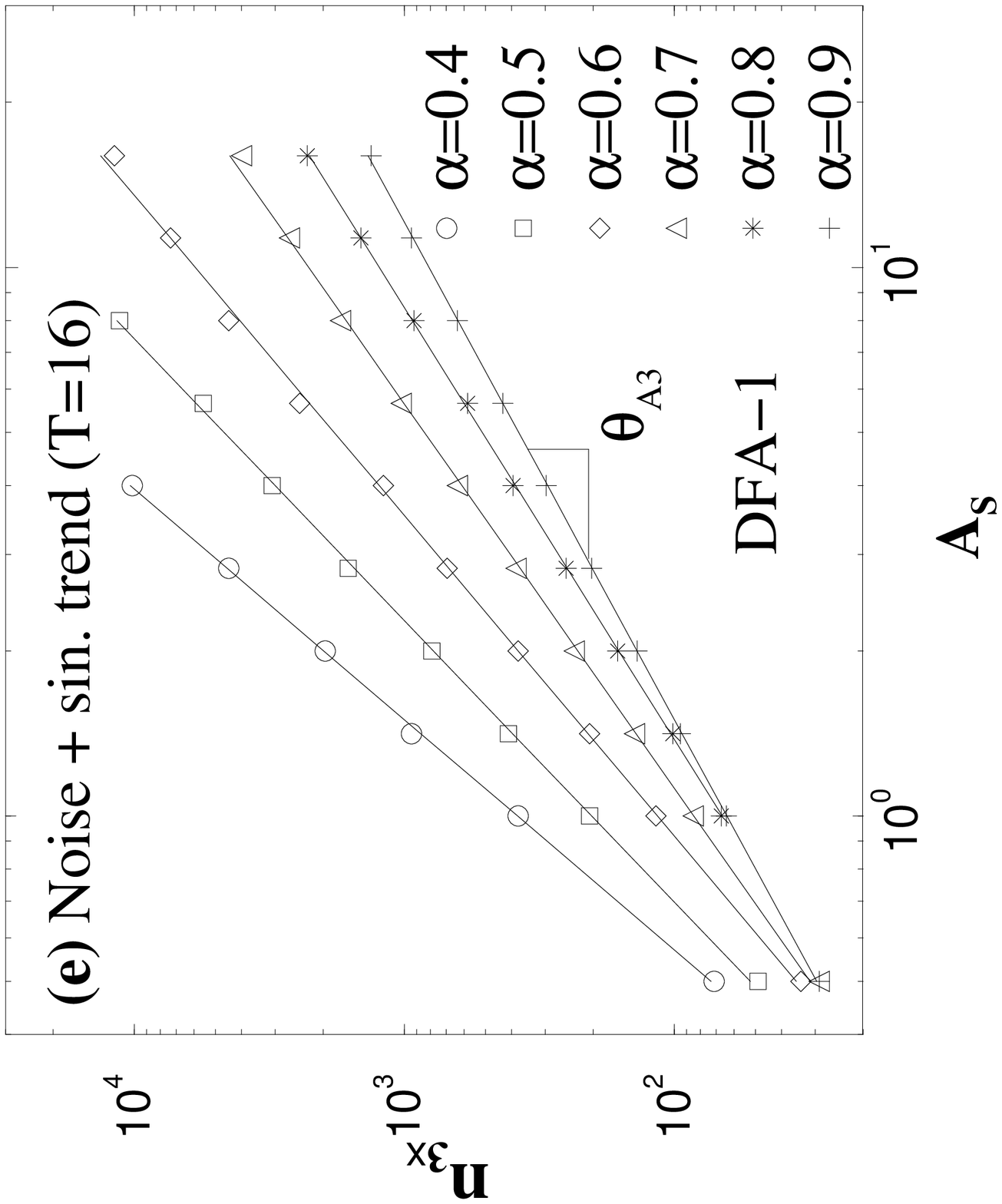}}}}
\caption{Dependence of the three crossovers in $F_{\rm \eta S}(n)$
for noise with a sinusoidal trend (Fig.~\ref{dfa1_nbw_n17}) on the
period $T$, and amplitude $A_{\rm S}$ of the sinusoidal trend. (a)
Power-law relation between the first crossover scale $n_{1\times}$
and the period $T$ for fixed amplitude $A_{\rm S}$ and varying
correlation exponent $\alpha$: $n_{1\times} \sim T^{\theta_{\rm T1}}$,
where $\theta_{\rm T1}$ is a positive crossover exponent [see Table
\ref{slope1} and Eq.~\ref{nx1}]. (b) Power-law relation between
the first crossover $n_{1\times}$ and the amplitude of the
sinusoidal trend $A_{\rm S}$ for fixed period $T$ and varying
correlation exponent $\alpha$: $n_{1\times} \sim A_{\rm S}^{\theta_{\rm
A1}}$ where $\theta_{\rm A1}$ is a negative crossover exponent [Table
\ref{slope1} and Eq.~(\ref{nx1})]. (c) The second crossover scale
$n_{2\times}$ depends only on the period $T$: $n_{2\times} \sim
T^{\theta_{\rm T2}}$, where $\theta_{\rm T2}\approx 1$. (d) Power-law
relation between the third crossover $n_{3\times}$ and $T$ for
fixed amplitude $A_{\rm S}$ and varying $\alpha$ trend:
$n_{3\times} \sim T^{\theta_{\rm T3} }$. (e) Power-law relation between
the third crossover $n_{3\times}$ and $A_{\rm S}$ for fixed $T$
and varying $\alpha$: $n_{3\times} \sim \left(A_{\rm
S}\right)^{\theta_{A3} }$. We find that $\theta_{\rm A3}=\theta_{\rm T3}$ [Table
\ref{slope3} and Eq.~(\ref{nx3})].} \label{S_dfa1_nbw_n17}
\end{figure}

First, we consider $n_{1\times}$. Surprisingly, we find that for
noise with any given correlation exponent $\alpha$ the crossover
scale $n_{1\times}$ exhibits long-range
power-law dependence of the period $T$ --- $n_{1\times} \sim
T^{\theta_{\rm T1}}$, and the amplitude $A_{\rm S}$ --- $n_{1\times}
\sim \left(A_{\rm S}\right)^{\theta_{\rm A1}}$ of the sinusoidal trend
[see Fig.~\ref{S_dfa1_nbw_n17}(a) and (b)]. We find that the
"crossover exponents" $\theta_{\rm T1}$ and $\theta_{\rm A1}$ have the same
magnitude but different sign --- $\theta_{\rm T1}$ is positive while
$\theta_{\rm A1}$ is negative. We also find that the magnitude of
$\theta_{\rm T1}$ and $\theta_{\rm A1}$ increases for the larger values of
the correlation exponents $\alpha$ of the noise. We present the
values of $\theta_{\rm T1}$ and $\theta_{\rm A1}$ for different correlation
exponent $\alpha$ in Table~\ref{slope1}. To understand these
power-law relations between $n_{1\times}$ and $T$, and  between
$n_{1\times}$ and $A_{\rm S}$, and also how the crossover scale
$n_{1\times}$ depends on the correlation exponent $\alpha$ we
employ the superposition rule [Eq.~\ref{addnp}] and estimate
$n_{1\times}$ analytically as the first intercept
$n_{1\times}^{th}$ of $F_{\rm \eta}(n)$ and $F_{\rm S}(n)$. From Eqs.~(\ref{eqpl}) and (\ref{dfa1_n}), we obtain the following
dependence of $n_{1\times}$ on $T$, $A_{\rm S}$ and $\alpha$:
\begin{equation}
n_{1\times} = \left(\frac{b_0}{k_1} \frac{T}{A_{\rm S}}\right)^{1/(2 - \alpha)}
\label{nx1}
\end{equation}
From this analytical calculation we obtain the following relation
between the two crossover exponents $\theta_{\rm T1}$ and $\theta_{\rm A1}$
and the correlation exponent $\alpha$: $\theta_{\rm T1}~=~-~\theta_{\rm A1}~=~1/{(2-\alpha)}$, which is in a good agreement with the
observed values of $\theta_{\rm T1}$, $\theta_{\rm A1}$ obtained from simulations
[see Fig.~\ref{S_dfa1_nbw_n17}(a) (b) and Table~\ref{slope1}].

Next, we consider $n_{2\times}$. Our analysis of the rms
fluctuation function $F_{\rm S}(n)$ for the sinusoidal signal in
Fig.~\ref{dfa1_puresin} suggests that the crossover scale $F_{\rm
S}(n)$ does not depend on the amplitude $A_{\rm S}$ of the
sinusoidal. The behavior of the rms fluctuation function
$F_{\rm \eta S}(n)$ for noise with superimposed sinusoidal trend
in Fig.~\ref{dfa1_nbw_n17}(a) and (b) indicates that $n_{2\times}$
does not depend on the correlation exponent $\alpha$ of the noise,
since for both correlated ($\alpha=0.9$) and anticorrelated
($\alpha=0$) noise ($T$ and $A_{\rm S}$ are fixed), the crossover
scale $n_{2\times}$ remains unchanged. We find that $n_{2\times}$
depends \textbf{only} on the period $T$ of the sinusoidal trend
and exhibits a long-range power-law behavior $n_{2\times} \sim
T^{\theta_{\rm T2}}$ with a crossover exponent $\theta_{\rm T2}\approx 1$
(Fig.~\ref{S_dfa1_nbw_n17}(c)) which is in agreement with the
prediction of Eq.(\ref{eqnx2}).

For the third crossover scale $n_{3\times}$, as for $n_{1\times}$
we find a power-law dependence on the period $T$, $n_{3\times}
\sim T^{\theta_{T3}}$, and amplitude $A_{\rm S}$, $n_{3\times} \sim
\left(A_{\rm S}\right)^{\theta_{A3}}$,of the sinusoidal trend [see
Fig.~\ref{S_dfa1_nbw_n17}(d) and (e)]. However, in contrast to the
$n_{1\times}$ case, we find that the crossover exponents $\theta_{\rm
Tp3}$ and $\theta_{\rm A3}$ are equal and positive with decreasing
values for increasing correlation exponents $\alpha$. In Table~\ref{slope3}, we present
the values of these two exponents for different correlation
exponent $\alpha$. To understand how the scale $n_{3\times}$
depends on $T$, $A_{\rm S}$ and the correlation exponent $\alpha$
simultaneously, we again employ the superposition rule
[Eq.~(\ref{addnp})] and estimate $n_{3\times}$ as the second
intercept $n_{3\times}^{th}$ of $F_{\rm \eta}(n)$ and $F_{\rm
S}(n)$. From Eqs.~(\ref{eqpp}) and (\ref{dfa1_n}), we obtain the
following dependence:
\begin{equation}
n_{3\times} = \left (\frac{1}{2 \sqrt{2} \pi b_0} A_{\rm S} T \right)^{1/\alpha}.
\label{nx3}
\end{equation}
From this analytical calculation we obtain $\theta_{\rm T3}=\theta_{\rm A3}
= 1/\alpha$ which is in good agreement with the values of $\theta_{\rm
T3}$ and $\theta_{\rm A3}$ observed from simulations
[Table~\ref{slope3}].

\begin{minipage}[t]{0.9\columnwidth}
\begin{table}[H!]
\caption{ The crossover exponents $\theta_{\rm T1}$ and $\theta_{\rm A1}$
characterizing the power-law dependence of $n_{1\times}$ on the
period $T$ and amplitude $A_{\rm S}$ obtained from simulations:
$n_{1\times} \sim T^{\theta_{\rm T1}}$ and $n_{1\times} \sim
\left(A_{\rm S}\right)^{\theta_{\rm A1}}$ for different value of the
correlation exponent $\alpha$ of noise
[Fig.~\ref{S_dfa1_nbw_n17}(a)(b)]. The values of $\theta_{\rm T1}$ and
$\theta_{\rm A1}$ are in good agreement with the analytical predictions
$\theta_{\rm T1}=-\theta_{\rm A1}=1/(2-\alpha)$
[Eq.~(\ref{nx1})].}\label{slope1}

\begin{tabular}{@{}c@{~~~~~~~~~~}c@{~~~~~~~~~}c@{~~~~~~}c@{}}
\centering $\alpha$ & $\theta_{\rm T1}$ & -$\theta_{\rm A1}$ & $1/(2-\alpha)$ \\
\tableline
  0.1 & 0.55  & 0.54 & 0.53\\
  0.3 & 0.58  & 0.59 & 0.59\\
  0.5 & 0.66  & 0.66 & 0.67\\
  0.7 & 0.74  & 0.75 & 0.77\\
  0.9 & 0.87  & 0.90 & 0.91\\
\end{tabular}
\end{table}
\end{minipage}
\hfill
\begin{minipage}[t]{0.9\columnwidth}
\begin{table}[H!]
\caption{The crossover exponents $\theta_{\rm T3}$ and $\theta_{\rm A3}$ for
the power-law relations: $n_{3\times} \sim T^{\theta_{\rm T3}}$ and
$n_{3\times} \sim \left(A_{\rm S}\right)^{\theta_{\rm A3}}$ for
different value of the correlation exponent $\alpha$ of noise
[Fig.~\ref{S_dfa1_nbw_n17}(c)(d)].  The values of $\theta_{p3}$ and
$\theta_{a3}$ obtained from simulations are in good agreement with the
analytical predictions $\theta_{\rm T3}=\theta_{\rm A3}=1/\alpha$
[Eq.~(\ref{nx3})].}

\begin{tabular}{@{}c@{~~~~~~~~~~}c@{~~~~~~~~~~~}c@{~~~~~~~~~~}c@{}}
\centering
$\alpha$ & $\theta_{\rm T3}$ & $\theta_{\rm A3}$ & $1/\alpha$\\ \tableline
  0.4 & 2.29 & 2.38 & 2.50\\
  0.5 & 1.92 & 1.95 & 2.00 \\
  0.6 & 1.69 & 1.71 & 1.67 \\
  0.7 & 1.39 & 1.43 & 1.43 \\
  0.8 & 1.26 & 1.27 & 1.25 \\
  0.9 & 1.06 & 1.10 & 1.11 \\
\end{tabular}
\label{slope3}
\hfill
\end{table}
\end{minipage}

Finally, our simulations show that all three crossover scales
$n_{1\times}$, $n_{2\times}$ and $n_{3\times}$ do not depend on
the length of the signal $N_{max}$, since $F_{\rm \eta}(n)$ and
$F_{\rm S}(n)$ do not depend on $N_{max}$ as shown in
Eqs.~(\ref{dfa1_n}), (\ref{addnp}), (\ref{eqpl}), and
(\ref{eqpp}).
\subsection{Higher order DFA on pure sinusoidal trend}\label{dfar_sin}

In the previous Sec.~\ref{secdfa1st}, we discussed how sinusoidal
trends affect the scaling behavior of correlated noise when the
DFA-1 method is applied. Since DFA-1 removes only constant trends in
data, it is natural to ask how the observed scaling results will
change when we apply DFA of order $\ell$ designed to remove
polynomial trends of order lower than $\ell$. In this section, we
first consider the rms fluctuation $F_{\rm S}$ for a sinusoidal
signal and then we study the scaling and crossover properties of
$F_{\rm \eta S}$ for correlated noise with superimposed sinusoidal
signal when higher order DFA is used.

\begin{figure}[H!]
\centerline{
\epsfysize=0.9\columnwidth{\rotate[r]{\epsfbox{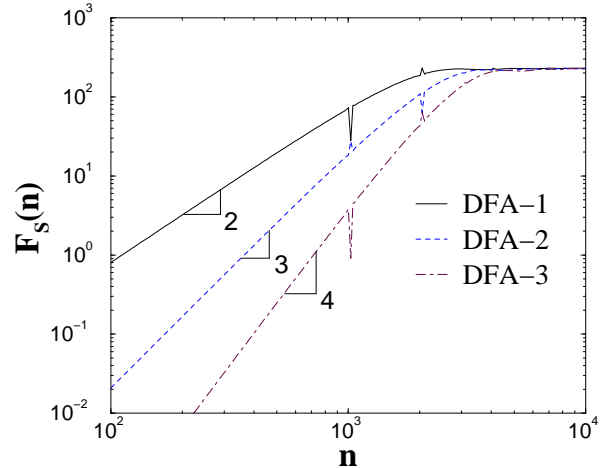}}}}
\vspace*{0.5cm} \caption{ Comparison of the results of different
order DFA on a sinusoidal trend.  The sinusoidal trend is
given by the function $64\sin({2\pi i}/2^{11})$ and the
length of the signal is $N_{max} = 2^{17}$. The spurious singularities (spikes) arise from the discrete data we use for the sinusoidal function.} \label{dfar}
\end{figure}

We find that the rms fluctuation function $F_{\rm S}$ does not
depend on the length of the signal $N_{max}$, and preserves a
similar shape when different order-$\ell$ DFA method is used
[Fig.~\ref{dfar}]. In particular, $F_{\rm S}$ exhibits a crossover
at a scale $n_{2\times}$ proportional to the period $T$ of the
sinusoidal: $n_{2\times} \sim T^{\theta_{\rm T2}}$ with $\theta_{\rm
T2}\approx 1 $. The crossover scale shifts to larger values for
higher order $\ell$ [Fig.~\ref{dfa1_puresin} and
Fig.~\ref{dfar}]. For the scale $n<n_{2\times}$, $F_{\rm S}$
exhibits an apparent scaling: $F_{\rm S}\sim n^{\alpha_{\rm S}}$
with an effective exponent $\alpha_{\rm S}~=~\ell +~1~$. For DFA-1, we
have $\ell=1$ and recover $\alpha_{\rm S}~=~2~$ as shown in
Eq.~(\ref{eqpl}). For $n>n_{2\times}$, $F_{\rm S}(n)$ is a
constant independent of the scale $n$, and of the order
$\ell$ of the DFA method in agreement with Eq.~(\ref{eqpp}).

Next, we consider $F_{\rm \eta S}(n)$ when DFA-$\ell$ with a higher order $\ell$ is used. We find that for all orders $\ell$, $F_{\rm \eta S}(n)$ does not depend on the length of the
signal $N_{max}$ and exhibits three crossovers --- at small,
intermediate and large scales --- similar behavior is reported for
DFA-1 in Fig.~\ref{dfa1_nbw_n17}. Since the crossover at small
scales, $n_{1\times}$, and the crossover at large scale,
$n_{3\times}$, result from the ``competition'' between the
scaling of the correlated noise and the effect of the sinusoidal
trend (Figs.~\ref{dfa1_nbw_n17} and \ref{x_nbw_2_128_n17}), using
the superposition rule [Eq.~(\ref{addnp})] we can estimate
$n_{1\times}$ and $n_{3\times}$ as the intercepts of $F_{\rm
\eta}(n)$ and $F_{\rm S}(n)$ for the general case of DFA-$\ell$.

For $n_{1\times}$ we find the following dependence on the period
$T$, amplitude $A_{\rm S}$, the correlation exponent $\alpha$ of
the noise, and the order $\ell$ of the DFA-$\ell$ method:
\begin{equation}
n_{1\times} \sim \left(T/A_{\rm S}\right )^{1/(\ell+1-\alpha)}
\label{nx1l}
\end{equation}
For DFA-1, we have $\ell=1$ and we recover Eq.~(\ref{nx1}). In
addition, $n_{1\times}$ is shifted to larger scales when higher
order DFA-$\ell$ is applied, due to the fact that the value of
$F_{\rm S}(n)$ decreases when $\ell$ increases ($\alpha_{\rm
S}=\ell+1$, see Fig.~\ref{dfar}).

For the third crossover observed in $F_{\rm \eta S}(n)$ at large
scale $n_{3\times}$ we find for all orders $\ell$ of the
DFA-$\ell$ the following scaling relation:
\begin{equation}
n_{3\times} \sim (T A_{\rm S})^{1/\alpha}\label{n3xl}.
\end{equation}
Since the scaling function $F_{\rm \eta}(n)$ for correlated noise
shifts vertically to lower values when higher order DFA-$\ell$
is used  [see the discussion in Appendix~\ref{secpuren} and
Sec.~\ref{secdfa12c}], $n_{3\times}$ exhibits a slight shift to
larger scales.

For the crossover $n_{2\times}$ in $F_{\rm \eta S}(n)$ at $F_{\rm
\eta S}(n)$ at intermediate scales, we find: $n_{2\times}~\sim~T$.
This relation is independent of the order $\ell$ of the DFA and is
identical to the relation found for $F_{\rm S}(n)$
[Eq.~(\ref{eqnx2})]. $n_{2\times}$ also exhibits a shift to larger
scales when higher order DFA is used [see Fig.~\ref{dfar}].

The reported here features of the crossovers in $F_{\rm \eta 
S}(n)$ can be used to identify low-frequency sinusoidal trends in
noisy data, and to recognize their effects on the scaling
properties of the data. This information may be useful when
quantifying correlation properties in data by means of scaling
analysis.

\section{Noise with Power-law trends} \label{seclc}

\begin{figure}[H!]
\centerline{
\epsfysize=0.9\columnwidth{\rotate[r]{\epsfbox{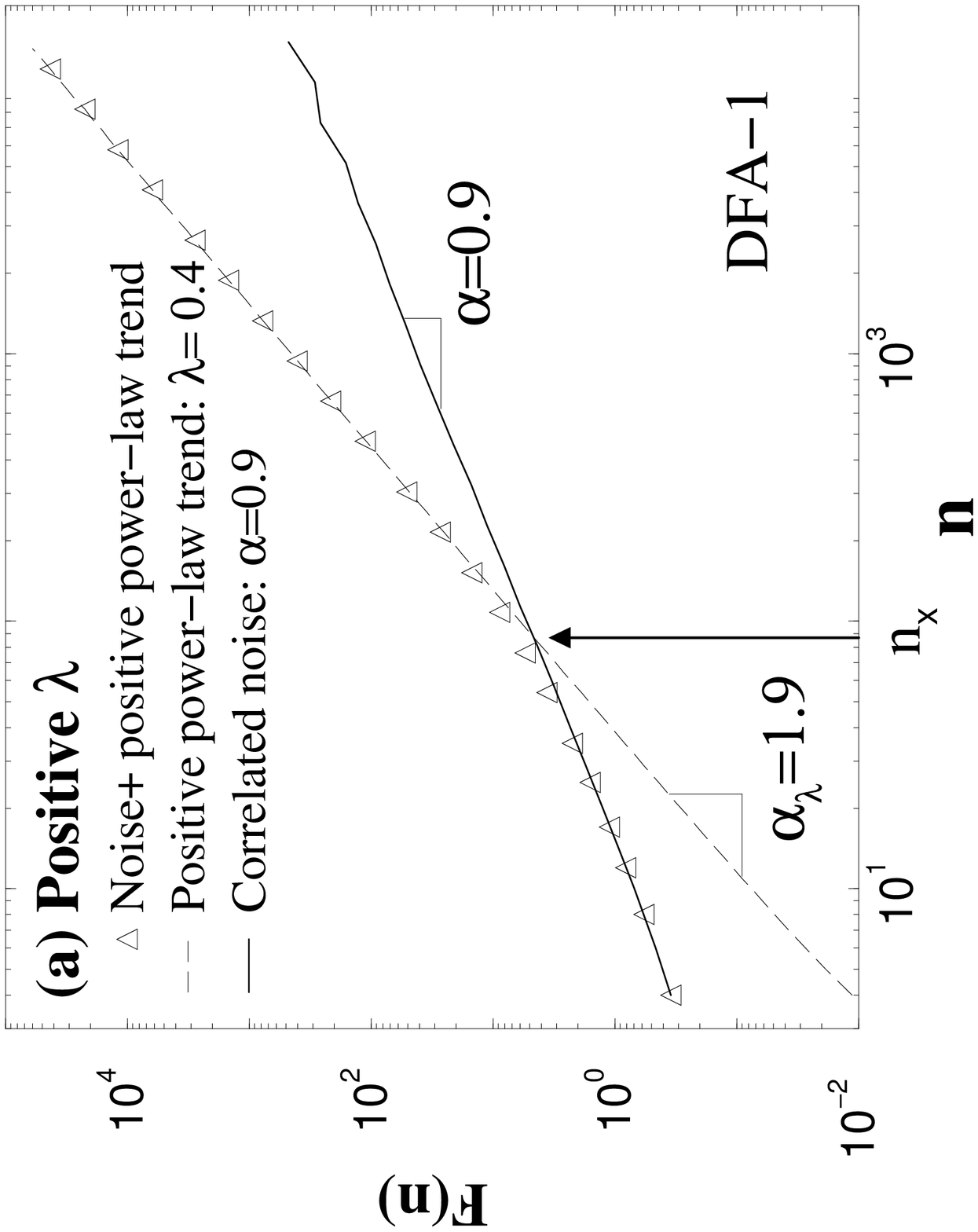}}}
}
\centerline{
\epsfysize=0.9\columnwidth{\rotate[r]{\epsfbox{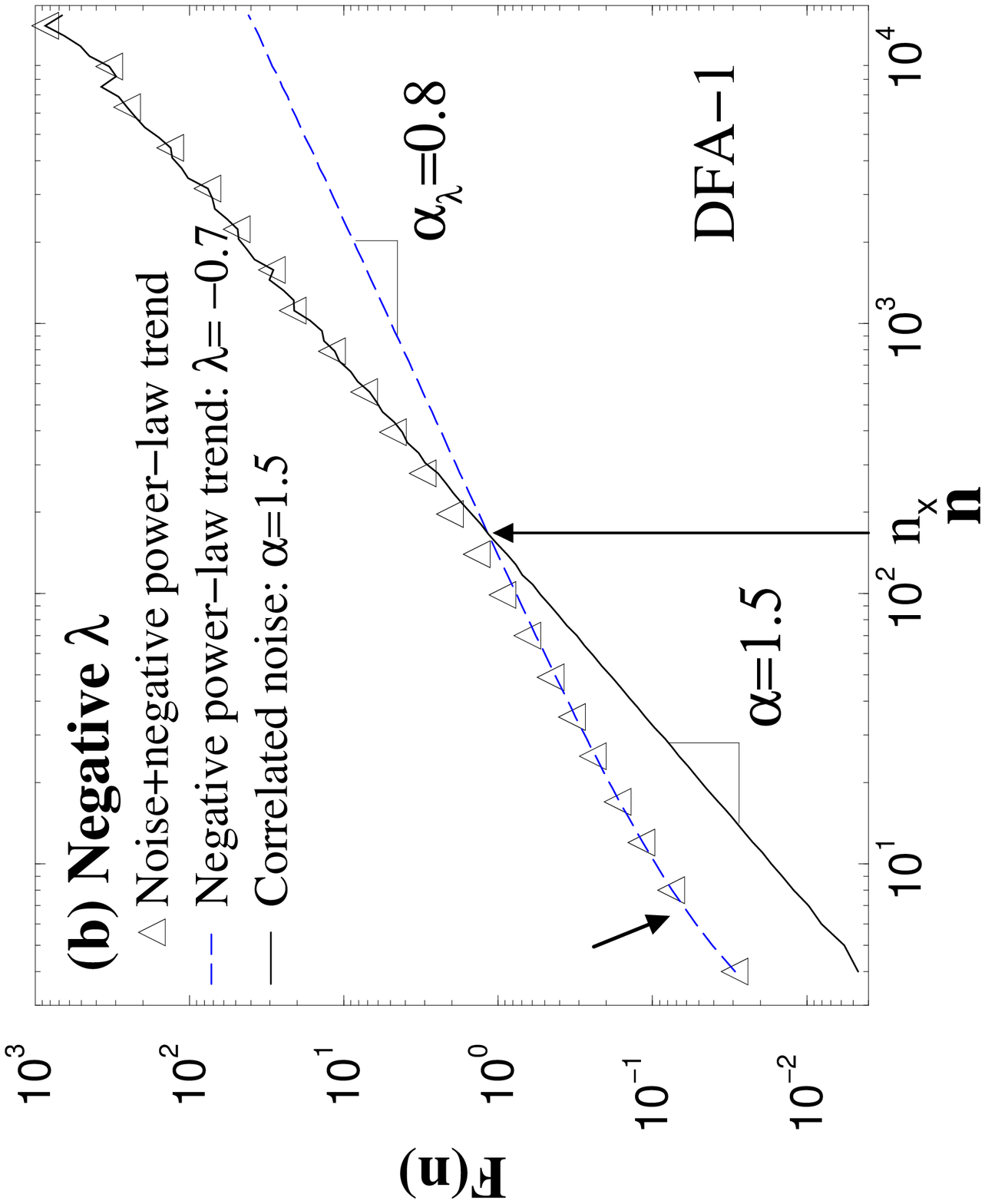}}}}
\vspace*{0.5cm}
\caption{Crossover behavior of the rms fluctuation function $F_{\rm \eta P}(n)$ (circles) for correlated noise (of length $N_{max}=2^{17}$) with a superimposed power-law trend $u(i)=A_{\rm P} i^{\lambda}$. The rms fluctuation function $F_{\rm \eta}(n)$ for noise (solid line) and the rms fluctuation function $F_{\rm P}(n)$ (dash line) are also shown for comparison. DFA-1 method is used. (a) $F_{\rm \eta P}(n)$ for noise with correlation exponent $\alpha_{\lambda}=0.9$, and power-law trend with amplitude $A_{\rm P} = 1000/ {\left( N_{max}\right)^{0.4}}$ and positive power $\lambda=0.4$; (b) $F_{\rm \eta P}(n)$ for Brownian noise (integrated white noise, $\alpha_{\lambda}=1.5$), and power-law trend with amplitude $A_{\rm P} = 0.01/\left(N_{max}\right)^{-0.7}$ and negative power $\lambda=-0.7$. Note, that although in both cases there is a ``similar'' crossover behavior for $F_{\rm \eta P}(n)$, the results in (a) and (b) represent completely opposite situations: while in (a) the power-law trend with positive power $\lambda$ dominates the scaling of $F_{\rm \eta P}(n)$ at large scales, in (b) the power-law trend with negative power $\lambda$ dominates the scaling at small scales, with arrow we indicate in (b) a weak crossover in $F_{\rm P}(n)$ (dashed lines) at small scales for negative power $\lambda$.}
\label{pn49.eps}
\end{figure}

In this section we study the effect of power-law trends on the
scaling properties of noisy signals. We consider the case of
correlated noise with superposed power-law trend $u(i)=A_{\rm P}
i^{\lambda}$, when $A_{\rm P}$ is a positive constant,
$i=1,...,N_{max}$, and $N_{max}$ is the length of the signal. We
find that when the DFA-1 method is used, the rms fluctuation
function $F_{\rm \eta P}(n)$ exhibits a crossover between two scaling regions
[Fig.~\ref{pn49.eps}]. This behavior results from the fact that at
different scales $n$, either the correlated noise or the power-law
trend is dominant, and can be predicted by employing the
superposition rule:
\begin{equation}
\left[F_{\rm \eta P}(n)\right]^2 = \left[F_{\rm \eta}(n)\right]^2 + \left[F_{\rm P}(n)\right]^2,
\label{eqaddp}
\end{equation}
where $F_{\rm \eta}(n)$ and $F_{\rm P}(n)$ are the rms fluctuation
function of noise and the power-law trend respectively, and $F_{\rm \eta P}(n)$ is the rms fluctuation function for the superposition of the noise and the power-law trend. Since the behavior of $F_{\rm \eta}(n)$ is known (Eq.~(\ref{dfa1_n}) and Appendix~\ref{secpuren}), we can understand the features of $F_{\rm \eta P}(n)$, if we know how $F_{\rm P}(n)$ depends on the characteristics of the power-law trend. We note that the scaling behavior of $F_{\rm \eta P}(n)$ displayed in Fig.~\ref{pn49.eps}(a) is to some extent similar to the behavior of the rms fluctuation function $F_{\rm \eta L}(n)$ for correlated noise with a linear trend [Fig.~\ref{dfa1_npbl_r_a01n17}] --- e.g. the noise is dominant at small scales $n$, while the trend is dominant at large scales. However, the behavior $F_{\rm P}(n)$ is more complex than that of $F_{\rm L}(n)$ for the linear trend, since the effective exponent $\alpha_{\lambda}$ for $F_{\rm P}(n)$ can depend on the power $\lambda$ of the power-law trend. In particular, for negative values of $\lambda$, $F_{\rm P}(n)$ can become dominated at small scales (Fig.~\ref{pn49.eps}(b)) while $F_{\rm \eta}(n)$ dominates at large scales --- a situation completely opposite of noise with linear trend (Fig.~\ref{dfa1_npbl_r_a01n17}) or with power-law trend with positive values for the power $\lambda$. Moreover, $F_{\rm P}(n)$ can exhibit crossover behavior at small scales [Fig.~\ref{pn49.eps}(b)] for negative $\lambda$ which is not observed for positive $\lambda$. In addition $F_{\rm P}(n)$ depends on the order $\ell$ of the DFA method and the length $N_{max}$ of the signal. We discuss the scaling features of the power-law trends in the following three subsections.

\subsection{Dependence of $F_{\rm P}(n)$ on the power $\lambda$}\label{secdfa1lc}
\begin{figure}[H!]
\centerline{
\epsfysize=0.47\textwidth{\rotate[r]{\epsfbox{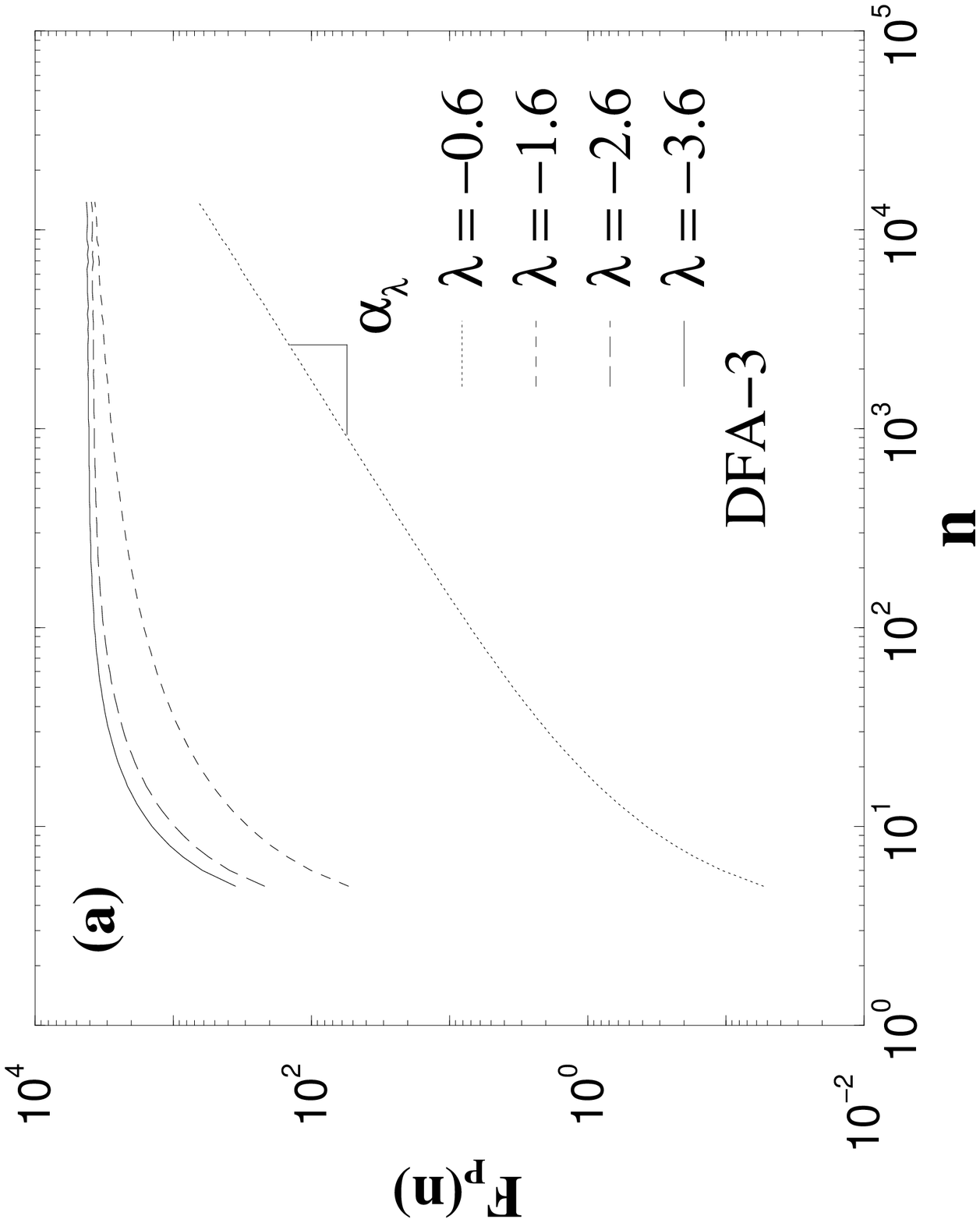}}}}
\centerline{
\epsfysize=0.45\textwidth{\rotate[r]{\epsfbox{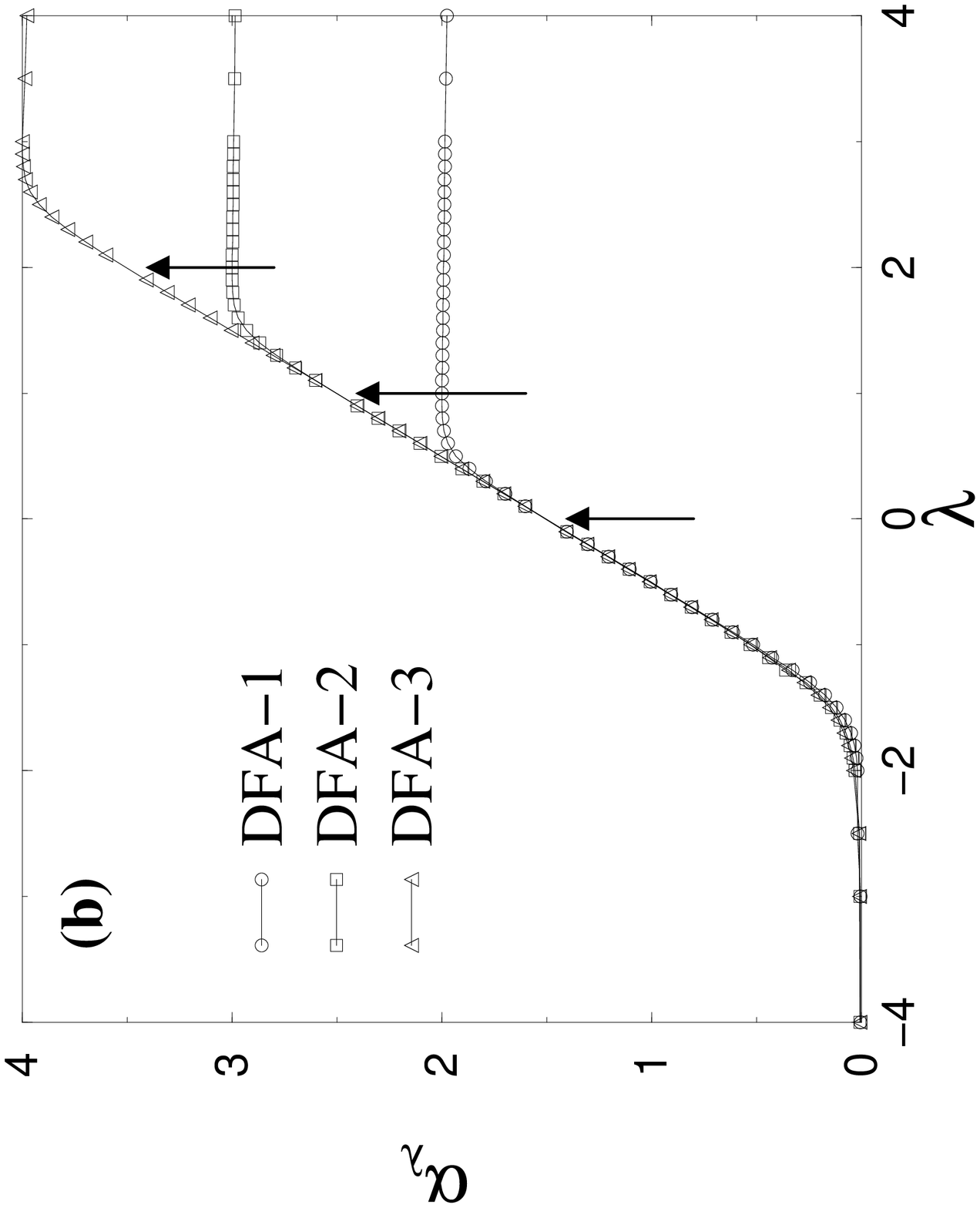}}}
}
\centerline{
\epsfysize=0.47\textwidth{\rotate[r]{\epsfbox{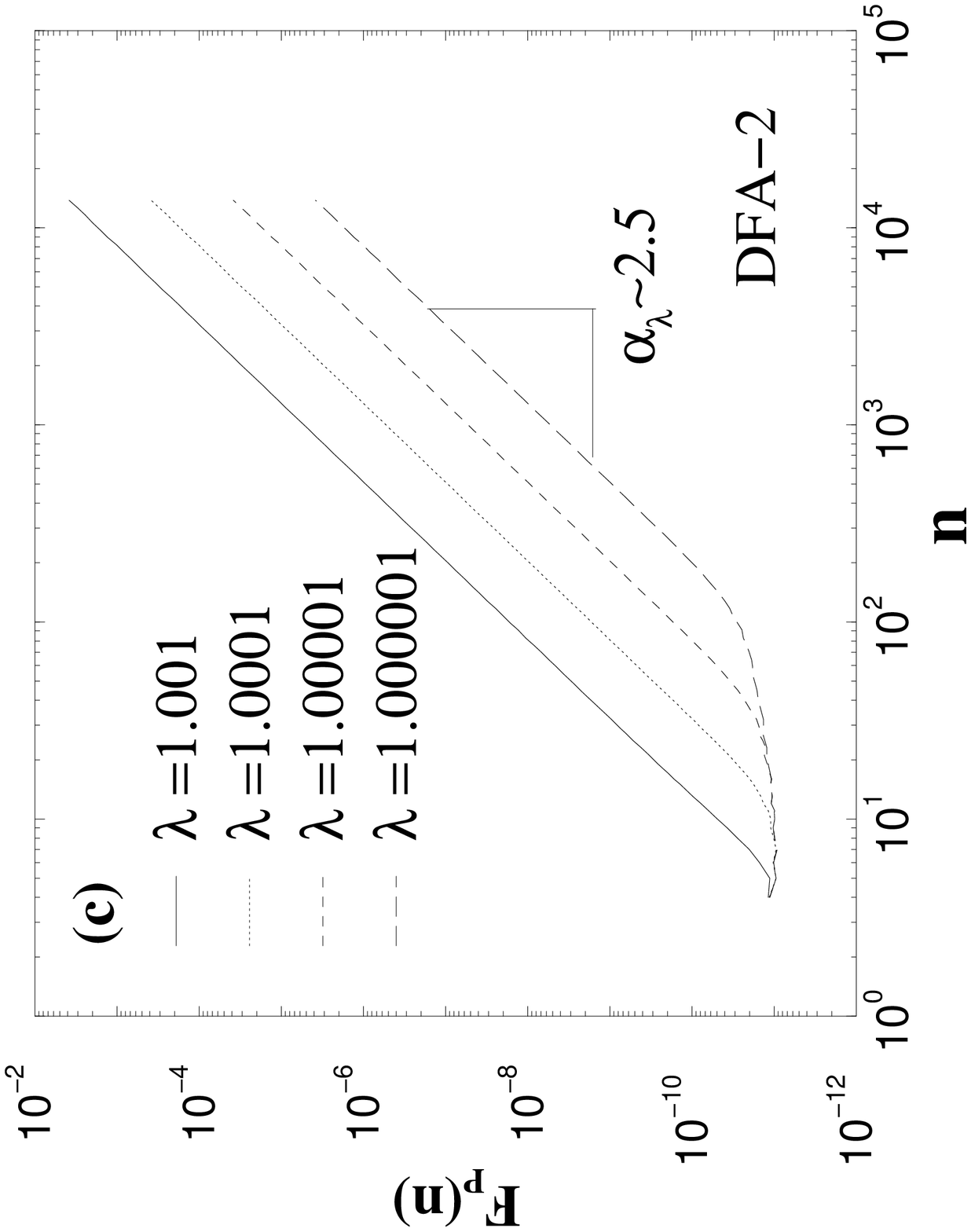}}}}
\vspace*{0.5cm}
\caption{Scaling behavior of rms fluctuation function $F_{\rm P}(n)$ for power-law trends, $u(i)\sim i^{\lambda}$, where $i=1,...,N_{max}$ and $N_{max}=2^{17}$ is the length of the signal. (a) For $\lambda<0$, $F_{\rm P}(n)$ exhibits crossover at small scales which is more pronounced with increasing the order $\ell$ of DFA-$\ell$ and decreasing the value of $\lambda$. Such crossover is not observed for $\lambda>0$ when $F_{\rm P}(n) \sim n^{\alpha_{\lambda}}$ for all scales $n$ [see Fig.~\ref{pn49.eps}(a)]. (b) Dependence of the effective exponent $\alpha_{\lambda}$ on the power $\lambda$ for different order $\ell=1,2,3$ of the DFA method. Three regions are observed depending on the order $\ell$ of the DFA: region I ($\lambda>\ell-0.5)$, where $\alpha_{\lambda} \approx \ell+1$; region II ($-1.5<\lambda<\ell-0.5$), where $\alpha_{\lambda}=\lambda+1.5$; region III ($\lambda<-1.5$), where $\alpha_{\lambda}\approx 0$. We note that for integer values of the power $\lambda=0,1,...,\ell-1$, where $\ell$ is the order of DFA we used, there is no scaling for $F_{\rm P}(n)$ and $\alpha_{\lambda}$ is not defined, as indicated by the arrows. (c) Asymptotic behavior near integer values of $\lambda$. $F_{\rm P}(n)$ is plotted for $\lambda \rightarrow 1$ when DFA-2 is used. Even for $\lambda-1 =10^{-6}$, we observe at large scales $n$ a region with an effective exponent $\alpha_{\lambda} \approx 2.5$, This region is shifted to infinitely large scales when $\lambda =1$.} \label{pslt.eps}
\end{figure}

First we study how the rms fluctuation function $F_{\rm P}(n)$ for a
power-law trend $u(i)=A_{\rm P} i^{\lambda}$ depends on the power $\lambda$. We find that 
\begin{equation}
F_{\rm P}(n) \sim A_{\rm P}n^{\alpha_{\lambda}},
\label{fplambda}
\end{equation} 
where $\alpha_{\lambda}$ is the effective exponent for the power-law trend. For positive $\lambda$ we observe no crossovers in $F_{\rm P}(n)$ (Fig.~\ref{pn49.eps}(a)). However, for negative $\lambda$ there is a crossover in $F_{\rm P}(n)$ at small scales $n$ (Fig.~\ref{pn49.eps}(b)), and we find that this crossover becomes even more pronounced with decreasing $\lambda$ or increasing the order $\ell$ of the DFA method, and is also shifted to larger scales [Fig.~\ref{pslt.eps}(a)].

Next, we study how the effective exponent $\alpha_{\lambda}$ for $F_{\rm P}(n)$ depends on the value of the power $\lambda$ for the power-law trend. We examine the scaling of $F_{\rm P}(n)$ and estimate $\alpha_{\lambda}$ for $-4<\lambda<4$. In the cases when $F_{\rm P}(n)$ exhibits a crossover, in order to obtain $\alpha_{\lambda}$ we fit the range of larger scales to the right of the crossover. We find that for any order $\ell$ of the DFA-$\ell$ method there are three regions with different relations between $\alpha_{\lambda}$ and $\lambda$ [Fig.~\ref{pslt.eps}(b)]: 
\begin{description}
\item (i) $\alpha_{\lambda}\approx \ell+1$ for $\lambda>\ell-0.5$ (region I); 
\item (ii) $\alpha_{\lambda} \approx \lambda +1.5$ for $-1.5\le \lambda \le \ell-0.5$ (region II); 
\item (iii) $\alpha_{\lambda} \approx 0$ for $\lambda<-1.5$ (region III). 
\end{description}
Note, that for integer values of the power $\lambda$ ($\lambda =0,1,...,m-1$), i.e. polynomial trends of order $m-1$, the DFA-$\ell$ method of order $\ell>m-1$ ($\ell$ is also an integer) leads to $F_{\rm P}(n) \approx 0$, since DFA-$\ell$ is designed to remove polynomial trends. Thus for a integer values of the power $\lambda$ there is no scaling and the effective exponent $\alpha_{\lambda}$ is not defined if a DFA-$\ell$ method of order $\ell > \lambda$ is used [Fig.~\ref{pslt.eps}]. However, it is of interest to examine the asymptotic behavior of the scaling of $F_{\rm P}(n)$ when the value of the power $\lambda$ is close to an integer. In particular , we consider how the scaling of $F_{\rm P}(n)$ obtained from DFA-2 method changes when $\lambda \rightarrow 1$ [Fig.~\ref{pslt.eps}(c)]. Surprisingly, we find that even though the values of $F_{\rm P}(n)$ are very small at large scales, there is a scaling for $F_{\rm P}(n)$ with a smooth convergence of the effective exponent $\alpha_{\lambda} \rightarrow 2.5$ when $\lambda \rightarrow 1$, according to the dependence $\alpha_{\lambda} \approx \lambda+1.5$ established for region II [Fig.~\ref{pslt.eps}(b)]. At smaller scales there is a flat region which is due to the fact that the fluctuation function $Y(i)$ (Eq.~(\ref{psi})) is smaller than the precision of the numerical simulation.

\subsection{Dependence of $F_{\rm P}(n)$ on the order $\ell$ of DFA}\label{secdfa12c}
Another factor that affects the rms fluctuation function of the power-law trend $F_{\rm P}(n)$, is the order $\ell$ of the DFA method used.  We first take into account that: 
\begin{description}
\item (1) for integer values of the power $\lambda$, the power-law trend $u(i)=A_{\rm P} i^{\lambda}$ is a polynomial trend which can be perfectly filtered out by the DFA method of order $\ell >\lambda$, and as discussed in Sec.~\ref{secdfa2bl} and Sec.~\ref{secdfa1lc} [see Fig.~\ref{pslt.eps}(b) and (c)], there is no scaling for $F_{\rm P}(n)$. Therefore, in this section we consider only non-integer values of $\lambda$. 
\item (2) for a given value of the power $\lambda$, the effective exponent $\alpha_{\lambda}$ can take different values depending on the order $\ell$ of the DFA method we use [see Fig.~\ref{pslt.eps}] --- e.g. for fixed $\lambda > \ell -0.5$, $\alpha_{\lambda} \approx \ell +1$. Therefore, in this section, we consider only the case when $\lambda<\ell -0.5$ (Region II and III).
\end{description}
\begin{figure}[H!]
\leftline{
\epsfysize=0.45\textwidth{\rotate[r]{\epsfbox{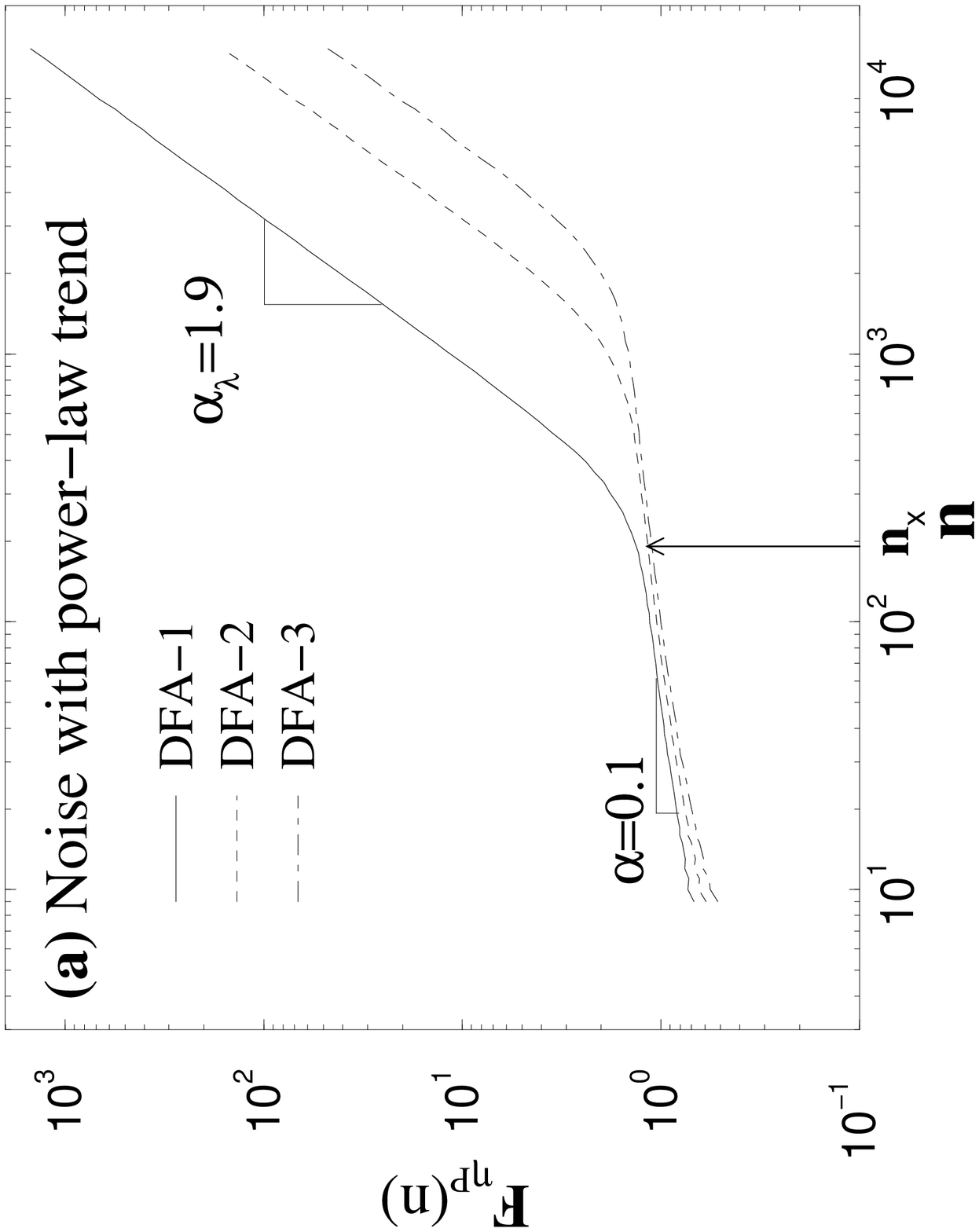}}}
}

\centerline{
\epsfysize=0.53\textwidth{\rotate[r]{\epsfbox{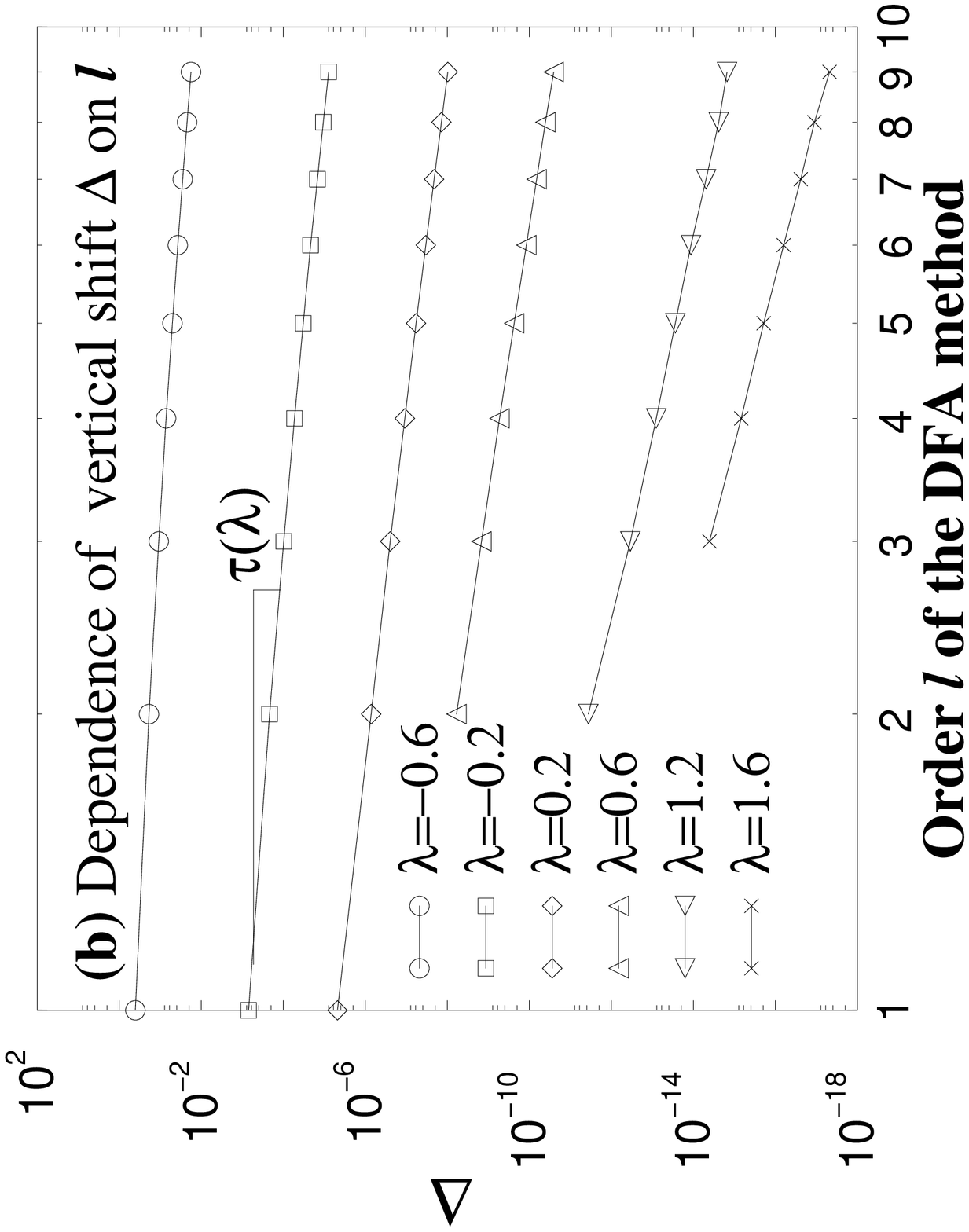}}}}

\centerline{
\epsfysize=0.45\textwidth{\rotate[r]{\epsfbox{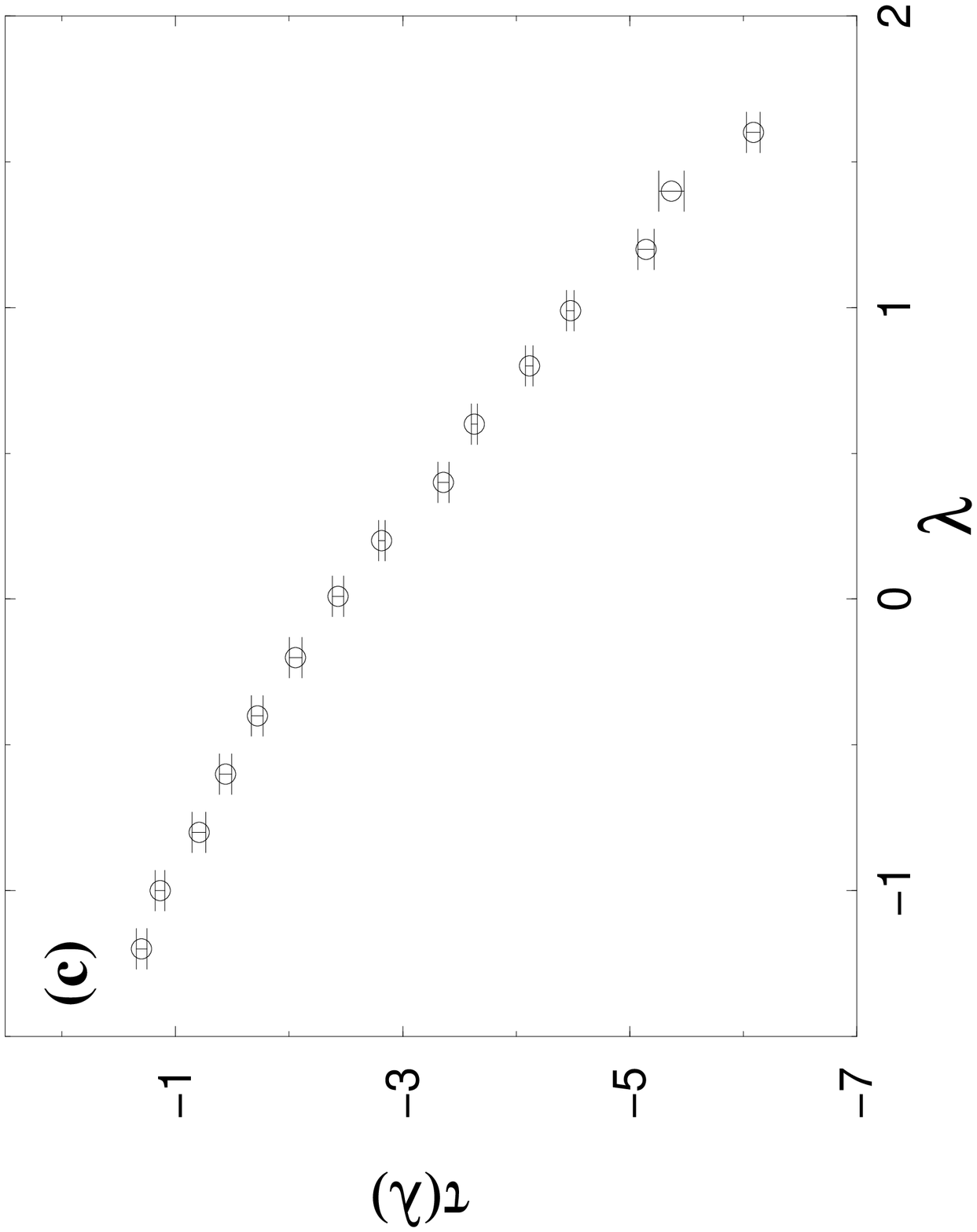}}}
}

\centerline{
\epsfysize=0.45\textwidth{\rotate[r]{\epsfbox{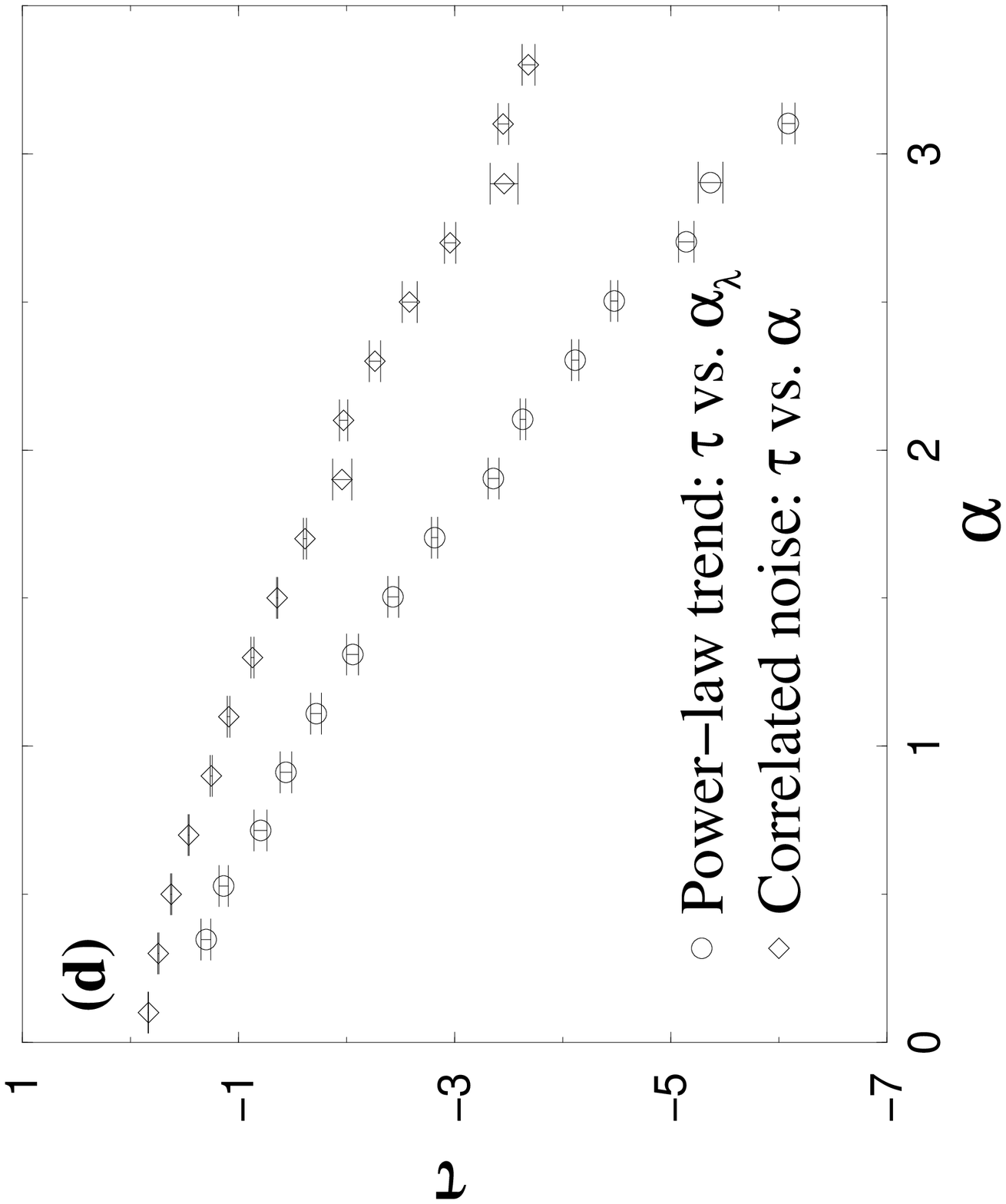}}}}

\caption{Effect of higher order DFA-$\ell$ on the rms fluctuation function $F_{\rm \eta P}(n)$ for correlated noise with superimposed power-law trend. (a) $F_{\rm \eta P}(n)$ for anticorrelated noise with correlation exponent $\alpha=0.1$ and a power-law $u(i)=A_{\rm P} i^{\lambda}$, where $A_{\rm P}=25/\left(N_{max}\right)^{0.4}$, $N_{max}=2^{17}$ and $\lambda=0.4$. Results for different order $\ell =1,2,3$ of the DFA method show (i) a clear crossover from a region at small scales where the noise dominates $F_{\rm \eta P}(n) \approx F_{\rm \eta}(n)$, to a region at larger scales where the power-law trend dominates $F_{\rm \eta P}(n) \approx F_{\rm P}(n)$, and (ii) a vertical shift $\Delta$ in $F_{\rm \eta P}$ with increasing $\ell$. (b) Dependence of the vertical shift $\Delta$ in the rms fluctuation function $F_{\rm P}(n)$ for power-law trend on the order $\ell$ of DFA-$\ell$ for different values of $\lambda$: $\Delta \sim \ell^{\tau(\lambda)}$. We define the vertical shift $\Delta$ as the y-intercept of $F_{\rm P}(n)$: $\Delta \equiv F_{\rm P}(n=1)$. Note, that we consider only non-integer values for $\lambda$ and that we consider the region $\lambda<\ell-0.5$. Thus, for all values of $\lambda$ the minimal order $\ell$ that can be used in the DFA method is $\ell >\lambda+0.5$. e.g. for $\lambda=1.6$ the minimal order of the DFA that can be used is $\ell=3$ (for details see Fig.~\ref{pslt.eps}(b)).
(c) Dependence of $\tau$ on the power $\lambda$ (error bars indicate the regression error for the fits of $\Delta(l)$ in (b)).
(d) Comparison of $\tau(\alpha_{\lambda})$ for $F_{\rm P}(n)$ and $\tau(\alpha)$ for $F_{\rm \eta}(n)$. Faster decay of $\tau(\alpha_{\lambda})$ indicates larger vertical shifts for $F_{\rm P}(n)$ compared to $F_{\rm \eta}(n)$ with increasing order $\ell$ of the DFA-$\ell$.} \label{dev_vs_order.eps}
\end{figure}

Since higher order DFA-$\ell$ provides a better fit for the data, the fluctuation function $Y(i)$ (Eq.~(\ref{psi})) decreases with increasing order $\ell$. This leads to a vertical shift to smaller values of the rms fluctuation function $F(n)$ (Eq.~(\ref{F})). Such a vertical shift is observed for the rms fluctuation function $F_{\rm \eta}(n)$ for correlated noise (see Appendix~\ref{secpuren}), as well as for the rms fluctuation function of power-law trend $F_{\rm P}(n)$. Here we ask how this vertical shift in $F_{\rm \eta}(n)$ and $F_{\rm P}(n)$ depends on the order $\ell$ of the DFA method, and if this shift has different properties for $F_{\rm \eta}(n)$ compared to $F_{\rm P}(n)$. This information can help identify power-law trends in noisy data, and can be used to differentiate crossovers separating scaling regions with different types of correlations, and crossovers which are due to effects of power-law trends.

We consider correlated noise with a superposed power-law trend,
where the crossover in $F_{\rm \eta P}(n)$ at large scales $n$
results from the dominant effect of the power-law trend ---
$F_{\rm \eta P}(n) \approx F_{\rm P}(n)$ (Eq.~(\ref{eqaddp}) and
Fig.~\ref{pn49.eps}(a)). We choose the power $\lambda<0.5$, a
range where for all orders $\ell$ of the DFA method the effective
exponent $\alpha_{\lambda}$ of $F_{\rm P}(n)$ remains the same ---
i.e. $\alpha_{\lambda}=\lambda+1.5$ (region II in
Fig.~\ref{pslt.eps}(b)). For a superposition of an anticorrelated
noise and power-law trend with $\lambda=0.4$, we observe
a crossover in the scaling behavior of $F_{\rm \eta P}(n)$, from a
scaling region characterized by the correlation exponent
$\alpha=0.1$ of the noise, where $F_{\rm \eta P}(n) \approx F_{\rm
\eta}(n)$, to a region characterized by an effective exponent
$\alpha_{\lambda}=1.9$, where $F_{\rm \eta P}(n) \approx F_{\rm
P}(n)$, for all orders $\ell=1,2,3$ of the DFA-$\ell$ method
[Fig.~\ref{dev_vs_order.eps}(a)].  We also find that the crossover
of $F_{\rm \eta P}(n)$ shifts to larger scales when the order
$\ell$ of DFA-$\ell$ increases, and that there is a vertical shift
of $F_{\rm \eta P}(n)$ to lower values. This vertical shift in
$F_{\rm \eta P}(n)$ at large scales, where $F_{\rm \eta P}(n)=F_{\rm P}(n)$, appears to be different in
magnitude when different order $\ell$ of the DFA-$\ell$ method is
used [Fig.~\ref{dev_vs_order.eps}(a)]. We also
observe a less pronounced vertical shift at small scales where
$F_{\rm \eta P}(n) \approx F_{\rm \eta}(n)$.

Next, we ask how these vertical shifts depend on
the order $\ell$ of DFA-$\ell$. We define the vertical shift $\Delta$ as the y-intercept of $F_{\rm P}(n)$: $\Delta \equiv F_{\rm P}(n=1)$. We find that the vertical shift $\Delta$ in $F_{\rm P}(n)$ for power-law trend follows a power law: $\Delta \sim \ell^{\tau(\lambda)}$. We tested this relation
for orders up to $\ell=10$, and we find that it holds for
different values of the power $\lambda$ of the power-law trend
[Fig.~\ref{dev_vs_order.eps}(b)]. Using Eq.~(\ref{fplambda}) we can write: $F_{\rm P}(n)/F_{\rm P}(n=1) = n^{\alpha_{\lambda}}$, i.e. $F_{\rm P}(n) \sim F_{\rm P}(n=1)$. Since $F_{\rm P}(n=1) \equiv \Delta \sim \ell^{\tau(\lambda)}$ [Fig.~\ref{dev_vs_order.eps}(b)], we find that:
\begin{equation}
F_{\rm P}(n) \sim \ell^{\tau(\lambda)}.
\label{fpl}
\end{equation}
We also find that the exponent
$\tau$ is negative and is a decreasing function of the power
$\lambda$ [Fig.~\ref{dev_vs_order.eps}(c)]. Because the effective
exponent $\alpha_{\lambda}$ which characterizes $F_{\rm P}(n)$
depends on the power $\lambda$ [see Fig.~\ref{pslt.eps}(b)], we
can express the exponent $\tau$ as a function of
$\alpha_{\lambda}$ as we show in Fig.~\ref{dev_vs_order.eps}(d).
This representation can help us compare the behavior of the
vertical shift $\Delta$ in $F_{\rm P}(n)$ with the shift in
$F_{\rm \eta}(n)$. For correlated noise with different correlation
exponent $\alpha$, we observe a similar power-law  relation
between the vertical shift in $F_{\rm \eta}(n)$ and the order
$\ell$ of DFA-$\ell$: $\Delta \sim \ell^{\tau(\alpha)}$, where
$\tau$ is also a negative exponent which decreases with $\alpha$.
In Fig.~\ref{dev_vs_order.eps}(d) we compare
$\tau(\alpha_{\lambda})$ for $F_{\rm P}(n)$ with $\tau(\alpha)$
for $F_{\rm \eta}(n)$, and find that for any
$\alpha_{\lambda}=\alpha$, $\tau(\alpha_{\lambda}) <
\tau(\alpha)$. This difference between the vertical shift for
correlated noise and for a power-law trend can be utilized to
recognize effects of power-law trends on the scaling properties of
data.

\subsection{Dependence of $F_{\rm P}(n)$ on the signal length $N_{max}$}\label{secdfa13c}

\begin{figure}[H!]
\leftline{
\epsfysize=0.44\textwidth{\rotate[r]{\epsfbox{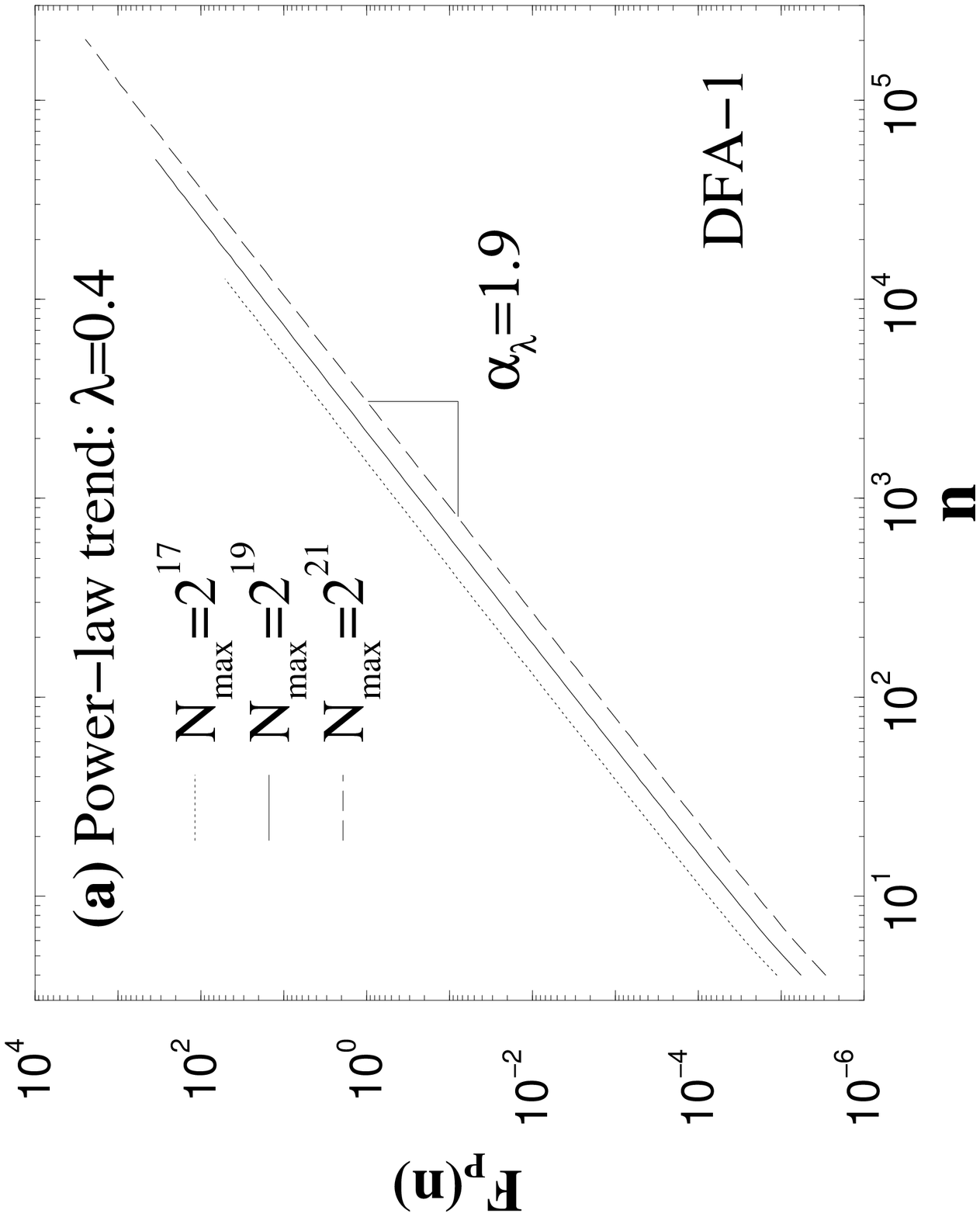}}}
}
\rightline{
\epsfysize=0.47\textwidth{\rotate[r]{\epsfbox{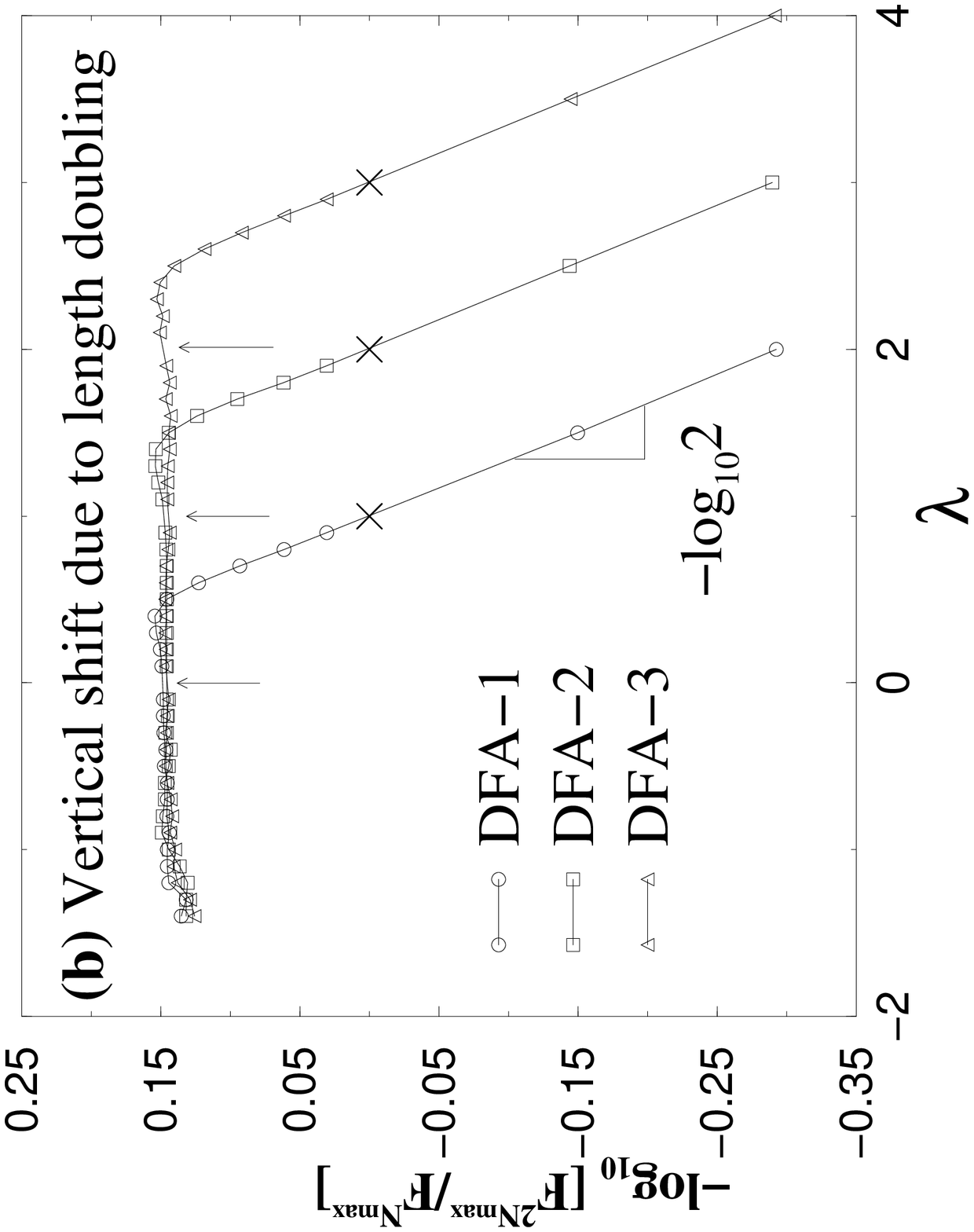}}}}
\caption{Dependence of the rms fluctuation function $F_{\rm P}(n)$ for power-law trend $u(i) = A_{\rm P} i^{\lambda}$, where $i=1,...,N_{max}$, on the length of the trend $N_{max}$. (a) A vertical shift is observed in $F_{\rm P}(n)$ for different values of $N_{max}$ --- $N_{1max}$ and $N_{2max}$. The figure shows that the vertical shift , defined as $F^{N_{1max}}_{\rm P}(n)/F^{N_{2max}}_{\rm P}(n)$, does not depend on $N_{max}$ but only on the ratio $N_{1max}/N_{2max}$, suggesting that $F_{\rm P}(n) \sim \left(N_{max}\right )^{\gamma}$. (b) Dependence of the vertical shift on the power $\lambda$. For $\lambda < \ell -0.5$ ($\ell$ is the order of DFA), we find a flat (constant) region characterized with effective exponent $\gamma =-0.5$ and negative vertical shift. For $\lambda > \ell -0.5$, we find an exponential dependence of the vertical shift on $\lambda$. In this region, $\gamma=\lambda-\ell$, and the vertical shift can be negative (if $\lambda<\ell$) or positive (if $\lambda>\ell$). the slope of $-\log_{10}\left(F^{2N_{max}}_{\rm P}(n)/F^{N_{max}}_{\rm P}(n)\right )$ vs. $\lambda$ is $-\log_{10}2$ due to doubling the length of the signal $N_{max}$. This slope changes to $-\log_{10}m$ when $N_{max}$ is increased $m$ times while $\gamma$ remains independent of $N_{max}$. For $\lambda=\ell$ there is no vertical shift, as marked with $\times$. Arrows indicate integer values of $\lambda<\ell$, for which values the DFA-$\ell$ method filters out completely the power-law trend and $F_{\rm P}=0$.}
\label{length.eps}
\end{figure}

Here, we study how the rms fluctuation function $F_{\rm P}(n)$ depends on the length $N_{max}$ of the power-law signal $u(i)=A_{\rm P} i^{\lambda}$ ($i=1,...,N_{max}$). We find that there is a vertical shift in $F_{\rm P}(n)$ with increasing $N_{max}$ [Fig.~\ref{length.eps}(a)]. We observe that when doubling the length $N_{max}$ of the signal the vertical shift in $F_{\rm P}(n)$, which we define as $F^{2N_{max}}_{\rm P}/F^{N_{max}}_{\rm P}$, remains the same, independent of the value of $N_{max}$. This suggests a power-law dependence of $F_{\rm P}(n)$ on the length of the signal:
\begin{equation}
F_{\rm P}(n) \sim \left(N_{max}\right )^{\gamma},
\label{fpnmax}
\end{equation}
where $\gamma$ is an effective scaling exponent. 

Next, we ask if the vertical shift depends on the power $\lambda$ of the power-law trend. When doubling the length $N_{max}$ of the signal, we find that for $\lambda < \ell -0.5$, where $\ell$ is the order of the DFA method, the vertical shift is a constant independent of $\lambda$ [Fig.~\ref{length.eps}(b)]. Since  the value of the vertical shift when doubling the length $N_{max}$ is $2^{\gamma}$ (from Eq.~(\ref{fpnmax})), the results in Fig.~\ref{length.eps}(b) show that $\gamma$ is independent of $\lambda$ when $\lambda < \ell-0.5$, and that $-\log 2^{\gamma} \approx -0.15$, i.e. the effective exponent $\gamma \approx -0.5$.

For $\lambda > \ell -0.5$, when doubling the length $N_{max}$ of the signal, we find that the vertical shift $2^{\gamma}$ exhibits the following dependence on $\lambda$:
$-\log _{10} 2^{\gamma} =\log _{10} 2^{\lambda-\ell}$, and thus the effective exponent $\gamma$ depends on $\lambda$ --- $\gamma = \lambda -\ell$. For positive integer values of $\lambda$ ($\lambda=\ell$), we find that $\gamma=0$, and there is no shift in $F_{\rm P}(n)$, suggesting that $F_{\rm P}(n)$ does not depend on the length $N_{max}$ of the signal, when DFA of order $\ell$ is used [Fig.~\ref{length.eps}]. Finally, we note that depending on the effective exponent $\gamma$, i.e. on the order $\ell$ of the DFA method and the value of the power $\lambda$, the vertical shift in the rms fluctuation function $F_{\rm P}(n)$ for power-law trend can be positive ($\lambda > \ell$), negative ($\lambda <\ell$), or zero ($\lambda=\ell$). 
 
\subsection{Combined effect on $F_{\rm P}(n)$ of $\lambda$, $\ell$ and $N_{max}$}

We have seen that, taking into account the effects of the power $\lambda$ (Eq.~(\ref{fplambda})), the order $\ell$ of DFA-$\ell$ (Eq.~(\ref{fpl})) and the effect of the length of the signal $N_{max}$ (Eq.~(\ref{fpnmax})), we reach the following expression for the rms fluctuation function $F_{\rm P}(n)$ for a power-law trend $u(i)=A_{\rm P} i^{\lambda}$:

\begin{eqnarray}
F_{\rm P}(n) \sim A_{\rm P} \cdot n^{\alpha_{\lambda}}\cdot \ell^{\tau(\lambda)}
\cdot \left(N_{max}\right )^{\gamma(\lambda)}, \label{eqntotal}
\end{eqnarray}
For correlated noise, the rms fluctuation function $F_{\rm \eta}(n)$ depends on the box size $n$ (Eq.~(\ref{dfa1_n})) and on the order $\ell$ of DFA-$\ell$ (Sec.~\ref{secdfa12c} and Fig.~\ref{dev_vs_order.eps}(a), (d)), and does not depend on the length of the signal $N_{max}$. Thus we have the following expression for $F_{\rm \eta}(n)$
\begin{equation}
F_{\rm \eta}(n) \sim n^{\alpha}\ell^{\tau(\alpha)},
\label{dfal_n}
\end{equation}

To estimate the crossover scale $n_{\times}$ observed in the apparent scaling of $F_{\rm \eta P}(n)$ for a correlated noise superposed with a power-law trend [Fig.~\ref{pn49.eps}(a), (b) and Fig.~\ref{dev_vs_order.eps}(a)], we employ the superposition rule (Eq.~(\ref{eqaddp})). From Eq.~(\ref{eqntotal}) and Eq.~(\ref{dfal_n}), we obtain $n_{\times}$ as the intercept between $F_{\rm P}(n)$ and $F_{\rm \eta}(n)$:
\begin{equation}
n_{\times} \sim \left[A l^{\tau(\lambda)-\tau(\alpha)} \left(N_{max}\right )^{\gamma}\right]^{1/(\alpha-\alpha_{\lambda})}.
\label{cspower}
\end{equation}
To test the validity of this result, we consider the case of correlated noise with a linear trend. For the case of a linear trend ($\lambda=1$) when DFA-1 ($\ell=1$) is applied, we have $\alpha_{\lambda}=2$ (see Appendix~\ref{secdfa1l} and Sec.~\ref{secdfa1lc}, Fig.~\ref{pslt.eps}(b)). Since in this case $\lambda = \ell =1 > \ell -0.5$ we have $\gamma =\lambda-\ell =0$ (see Sec.\ref{secdfa13c} Fig.~\ref{length.eps}(b)), and from Eq.~(\ref{cspower}) we recover Eq.~(\ref{si_bl}).

\section{Conclusion and Summary}\label{seccon}
In this paper we show that the DFA method performs better than the standard R/S analysis to quantify the scaling behavior of noisy signals for a wide range of correlations, and we estimate the range of scales where the performance of the DFA method is optimal. We consider different types of trends superposed on correlated noise, and study how these trends affect the scaling behavior of the noise. We demonstrate that there is a competition between a trend and
a noise, and that this competition can lead to crossovers in the scaling. We investigate the features of these crossovers, their dependence on the properties of the noise and the superposed trend. Surprisingly, we find that crossovers which are a result of trends can exhibit power-law dependences on the parameters of the trends. We show that these crossover phenomena can be explained by the superposition of the separate results of the DFA method on the noise and on the trend, assuming that the noise and the trend are not correlated, and that the scaling properties of the noise and the apparent scaling behavior of the trend are known. Our work may provide some help to differentiate between different types of crossovers --- e.g. crossovers which separate scaling regions with different correlation properties may differ from crossovers which are an artifact of trends. The results we present here could be useful for identifying the presence of trends and to accurately interpret correlation properties of noisy data.

\acknowledgments We thank NIH/National Center for Research
Resources (P41RR13622), NSF and the Spanish Government
(BIO99-0651-CO2-01) for support, and C.-K. Peng, A.L. Goldberger and Y. Ashkenazy for helpful
discussions. When concluding our work, we became aware of an independent study by J.W. Kantelhardt et. al\cite{Kantelhardtpha2001}, where similar issues are discussed. We thank J.W. Kantelhardt and A. Bunde for sending us their preprint before publication. 

\appendix

\section{Noise}\label{secpuren}

The standard signals we generate in our study are uncorrelated,
correlated, and anticorrelated noise. First we must have a clear idea
of the scaling behaviors of these standard signals before we use them to
study the effects from other aspects. We generate noises by using a
modified Fourier filtering method\cite{MFFM}. This method can
efficiently generate noise, $u(i)$ ($i=1,2,3,...,N_{\mbox{\scriptsize
max}}$), with the desired power-law correlation function which
asymptotically behaves as: $<|\sum\limits_{j=i}^{i+t}u(j)|^2> \sim
t^{2\alpha}$.  By default, a generated noise has standard deviation
$\sigma=1$.  Then we can test DFA and R/S by applying it on generated
noises since we know the expected scaling exponent $\alpha$.

\begin{figure}[H!]
\centerline{
\epsfysize=0.47\textwidth{\rotate[r]{\epsfbox{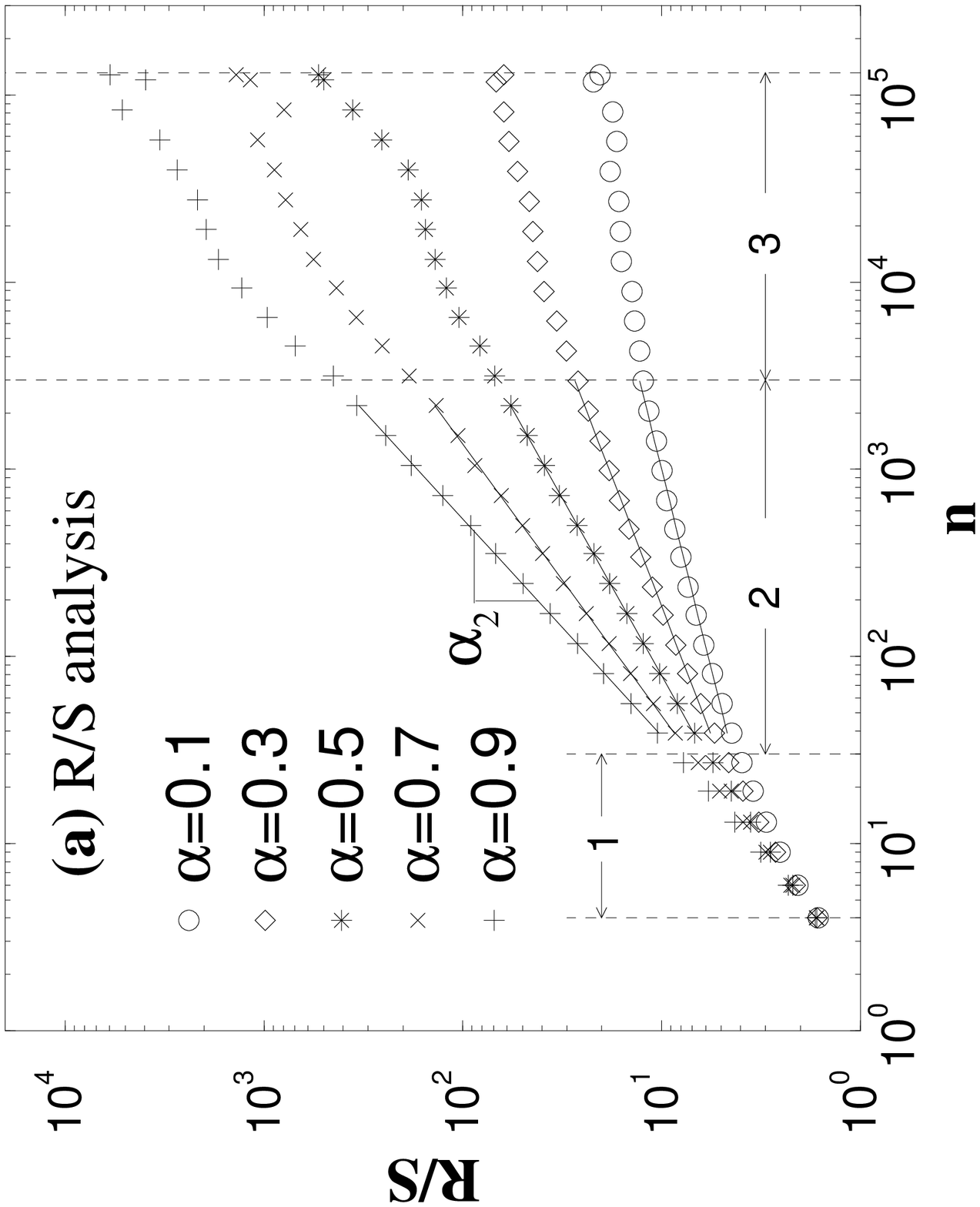}}}
}
\centerline{
\epsfysize=0.47\textwidth{\rotate[r]{\epsfbox{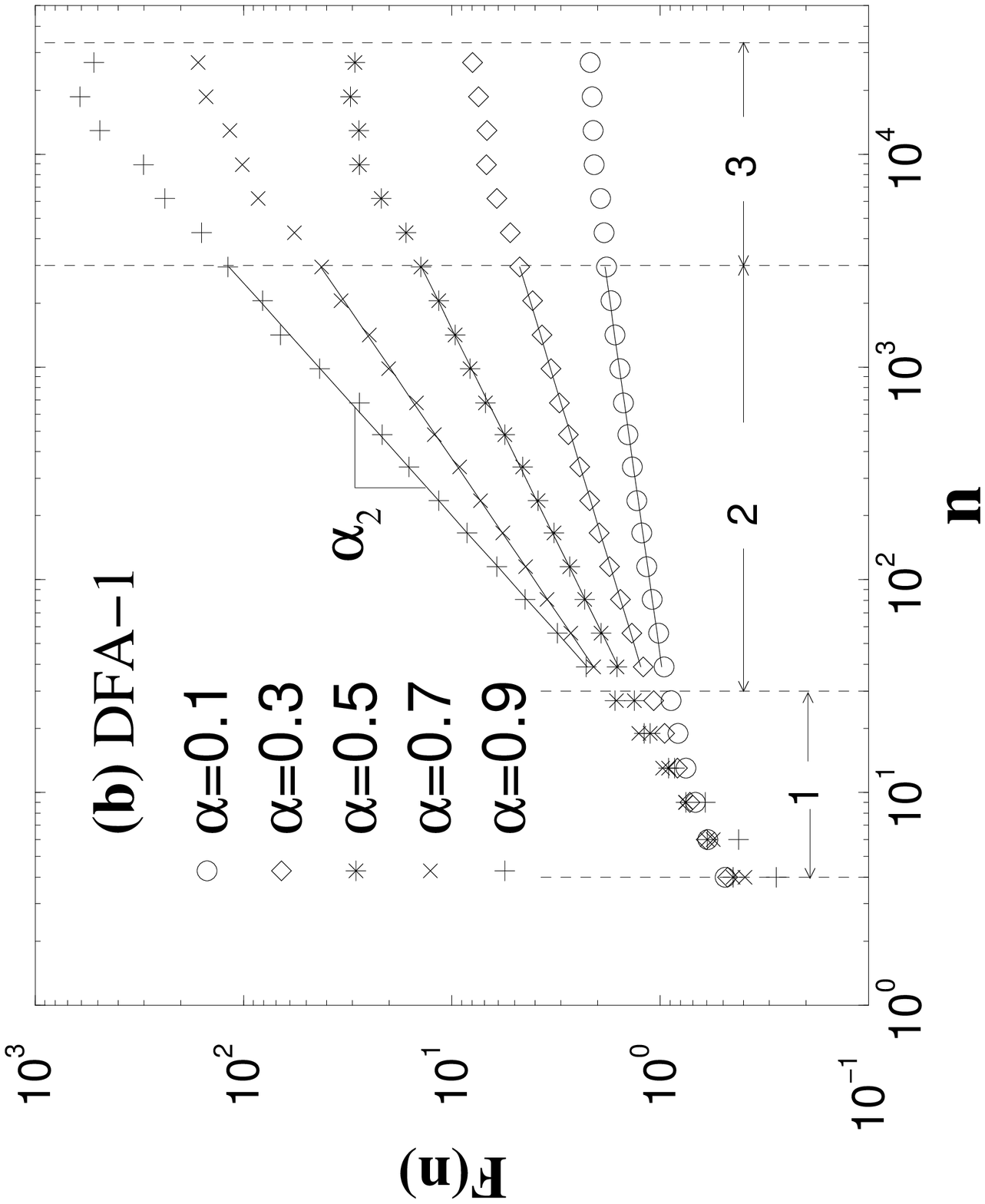}}}
}
\centerline{
\epsfysize=0.47\textwidth{\rotate[r]{\epsfbox{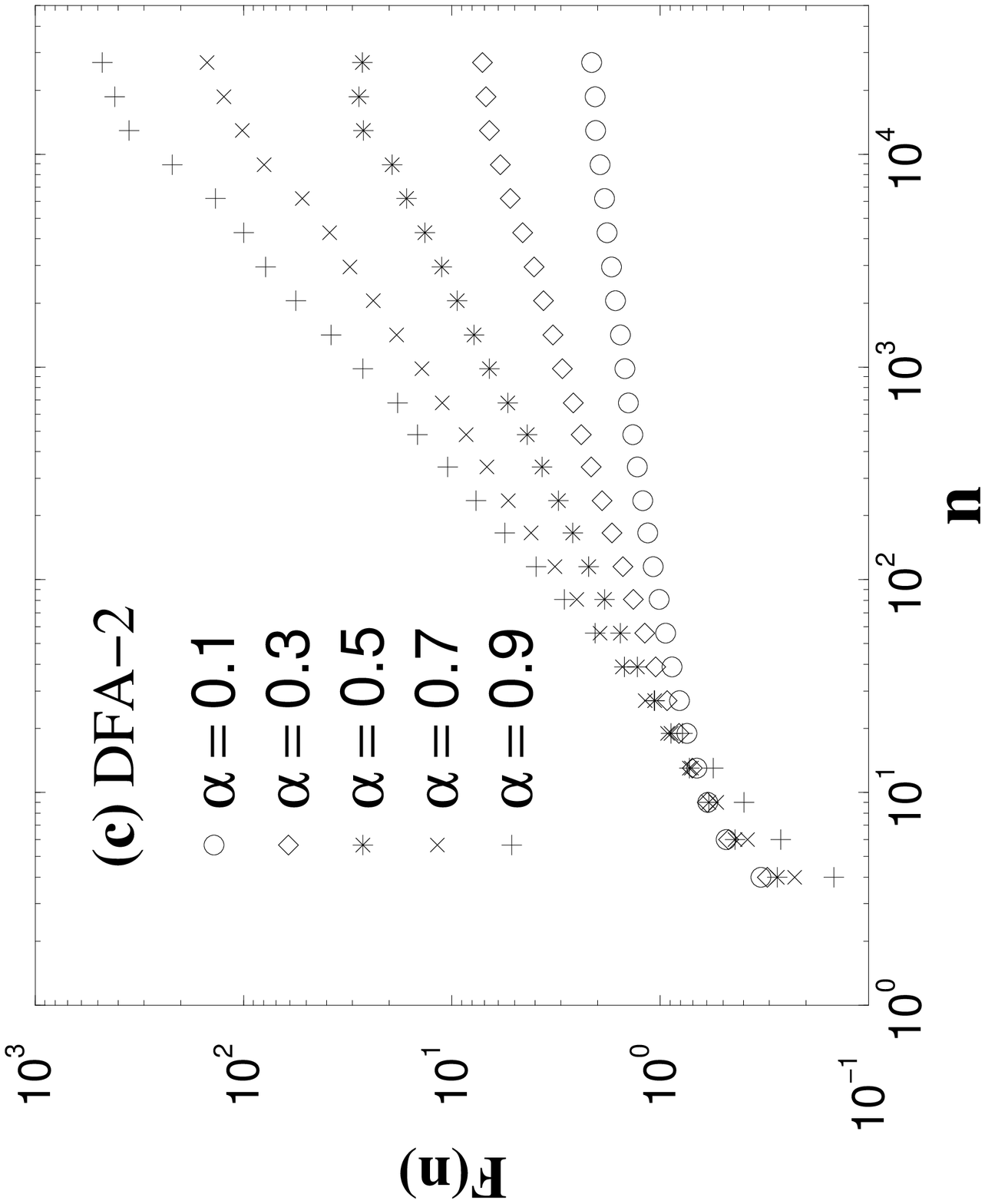}}}
}
\caption{Scaling behavior of noise with the scaling exponent
$\alpha$. The length of noise $N_{\mbox{\scriptsize max}}=2^{17}$.  (a)
Rescaled range analysis (R/S) (b) Order 1 detrended fluctuation analysis
(DFA-1) (c) Order 2 detrended fluctuation analysis.  We do the linear
fitting for R/S analysis and DFA-1 in three regions as shown and get
$\alpha_1$, $\alpha_2$ and $\alpha_3$ for estimated $\alpha$, which are
listed in the Table.\ref{tahaFT} and Table.\ref{tadfa1FT}. We find that
the estimation of $\alpha$ is different in the different region.}
\label{technique}
\end{figure}

Before doing that, we want to briefly review the algorithm of R/S
analysis. For a signal $u(i)$($i=1,...,N_{\mbox{\scriptsize
max}}$), it is divided into boxes of equal size $n$. In each box,
the \textit{cumulative departure}, $X_i$ (for $k$-th box,
$i=kn+1,..., kn+n$), is calculated
\begin{equation}
X_i =\sum\limits_{j=kn+1}^{i} (u(j)-<u>)
\end{equation}
where $<u>=n^{-1}\sum\limits_{i=kn+1}^{(k+1)n}u(i) $
, and the \textit{rescaled range} $R/S$ is defined by
\begin{equation}
R/S = S^{-1} \left[\max \limits_{kn+1\leq i \leq (k+1)n}X_i - \min\limits_{kn+1\leq i \leq
(k+1)n}X_i\right],
\end{equation}
where $S=
\sqrt{{n^{-1}}\sum\limits_{j=1}^{n}(u(j)-<u>)^2}$ is the standard deviation in each box.
The average of rescaled range in all the boxes of equal size $n$, is obtained and denoted by $<R/S>$.
Repeat the above computation over different box size $n$ to provide a relationship between $<R/S>$ and $n$. According to Hurst's experimental
study\cite{hurstanalysis}, a power-law relation between $<R/S>$ and the box size $n$ indicates the presence of scaling: $<R/S> \sim n^{\alpha}$.

Figure~\ref{technique} shows the results of R/S, DFA-1 and DFA-2 on the
same generated noises. Loosely speaking, we can see that $F(n)$ (for
DFA) and $R/S$ (for R/S analysis) show power-law relation with $n$ as
expected: $F(n) \sim n^{\alpha}$ and $R/S \sim n^{\alpha}$. In addition,
there is no significant difference between the results of different
order DFA except for some vertical shift of the curves and the little
bend-down for small box size $n$.  The bent-down for very small box of
$F(n)$ from higher order DFA is because there are more variables to fit
those few points.

\begin{minipage}[t]{0.9\columnwidth}
\begin{table}[H!]
\caption{ Estimated $\alpha$ of correlation noise from R/S analysis in
three regions as shown in Fig.\ref{technique}(a). $\alpha$ is the input
value of the scaling exponent, $\alpha_1$ is the estimated in the region
1 for $4<n\leq32$, $\alpha_2$ in the region 2 for $32<n\leq 3162$ and
$\alpha_3$ in the region 3 for $3126<n \leq 2^{17}$. Noise are the same
as used in Table.\ref{tadfa1FT}.} \label{tahaFT}

\begin{tabular}{@{}c@{~~~~~~~~}c@{~~~~~~~~}c@{~~~~~~~~}c@{}}
\centering
  $\alpha$ & $\alpha_1$ & $\alpha_2$  & $\alpha_3$  \\ \tableline
  0.1 & 0.44 & 0.23 & 0.12 \\
  0.3 & 0.52 & 0.37 & 0.23 \\
  0.5 & 0.62 & 0.52 & 0.47 \\
  0.7 & 0.72 & 0.70 & 0.45 \\
  0.9 & 0.81 & 0.87 & 0.63 \\
\end{tabular}
\end{table}
\end{minipage}
\begin{minipage}[t]{0.9\columnwidth}
\begin{table}
\caption{ Estimated $\alpha$ of correlation noise from DFA-1 in
three regions as shown in Fig.\ref{technique}(b). $\alpha$ is the
input value of the scaling exponent, $\alpha1$ is the estimated in
the region 1 for $4<n\leq32$, $\alpha2$ in the region 2 for
$32<n\leq 3162$ and $\alpha3$ in the region 3 for $3126<n \leq
2^{17}$.}\label{tadfa1FT}

\begin{tabular}{@{}c@{~~~~~~~~}c@{~~~~~~~~}c@{~~~~~~~~}c@{}}
\centering
$\alpha$ & $\alpha_1$  & $\alpha_2$ & $\alpha_3$ \\ \tableline
  0.1 & 0.28 & 0.15 & 0.08 \\
  0.3 & 0.40 & 0.31 & 0.22 \\
  0.5 & 0.55 & 0.50 & 0.35 \\
  0.7 & 0.72 & 0.69 & 0.55 \\
  0.9 & 0.91 & 0.91 & 0.69 \\
\end{tabular}
\end{table}
\end{minipage}

Ideally, when analyzing a standard noise, $F(n)$ (DFA) and $R/S$ ($R/S$
analysis) will be a power-law function with a given power: $\alpha$, no
matter which region of $F(n)$ and $R/S$ is chosen for calculation.
However, a careful study shows that the scaling exponent $\alpha$ depends
on scale $n$. The estimated $\alpha$ is different for the different
regions of $F(n)$ and $R/S$ as illustrated by Figs.~\ref{technique}(a)
and \ref{technique}(b) and by Tables~\ref{tahaFT} and \ref{tadfa1FT}.
It is very important to know the best fitting region of DFA and R/S
analysis in the study of real signals. Otherwise, the wrong $\alpha$
will be obtained if an inappropriate region is selected.

\begin{figure}[H!]
\centerline{
\epsfysize=0.47\textwidth{\rotate[r]{\epsfbox{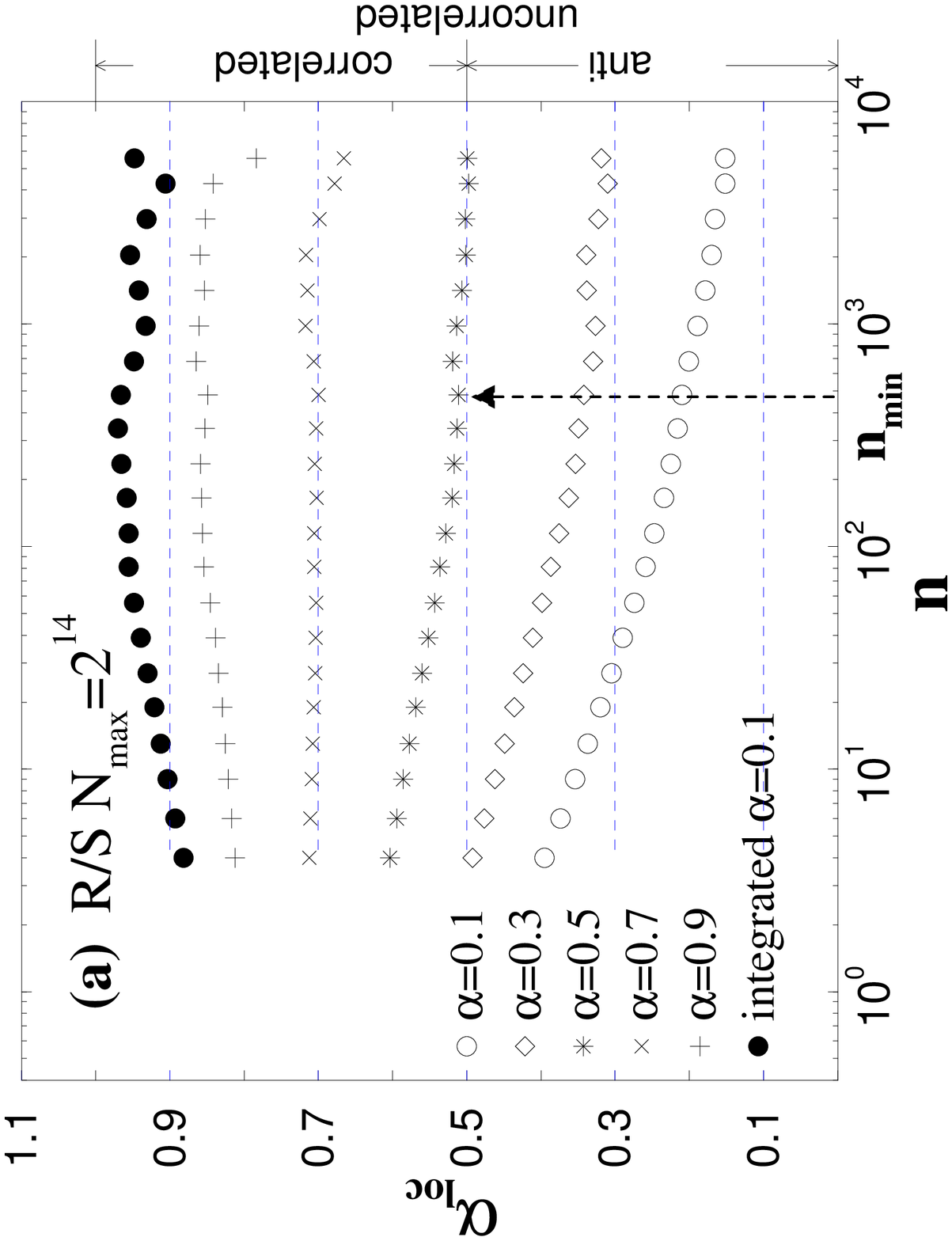}}}
}
\centerline{
\epsfysize=0.47\textwidth{\rotate[r]{\epsfbox{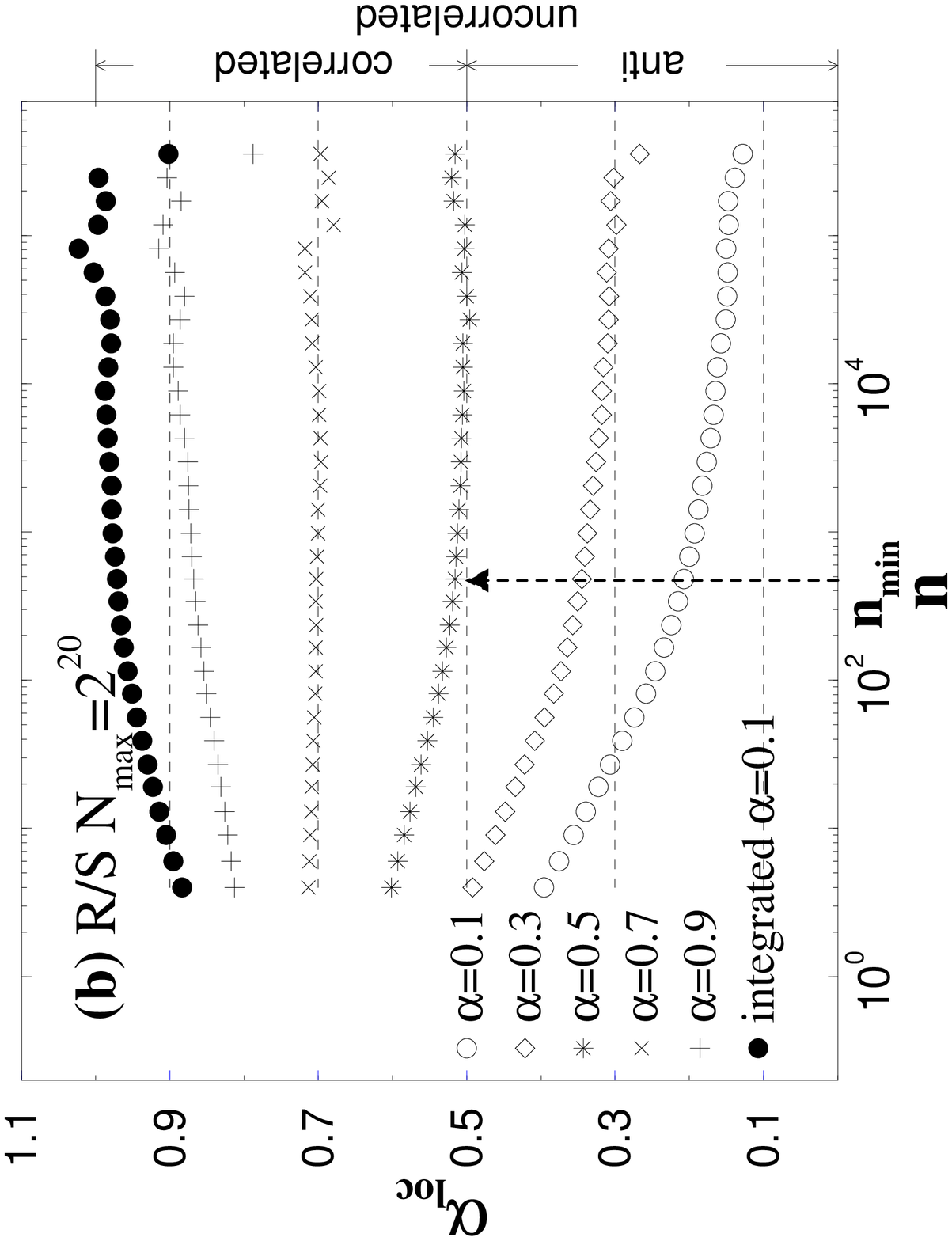}}}
}
\centerline{
\epsfysize=0.47\textwidth{\rotate[r]{\epsfbox{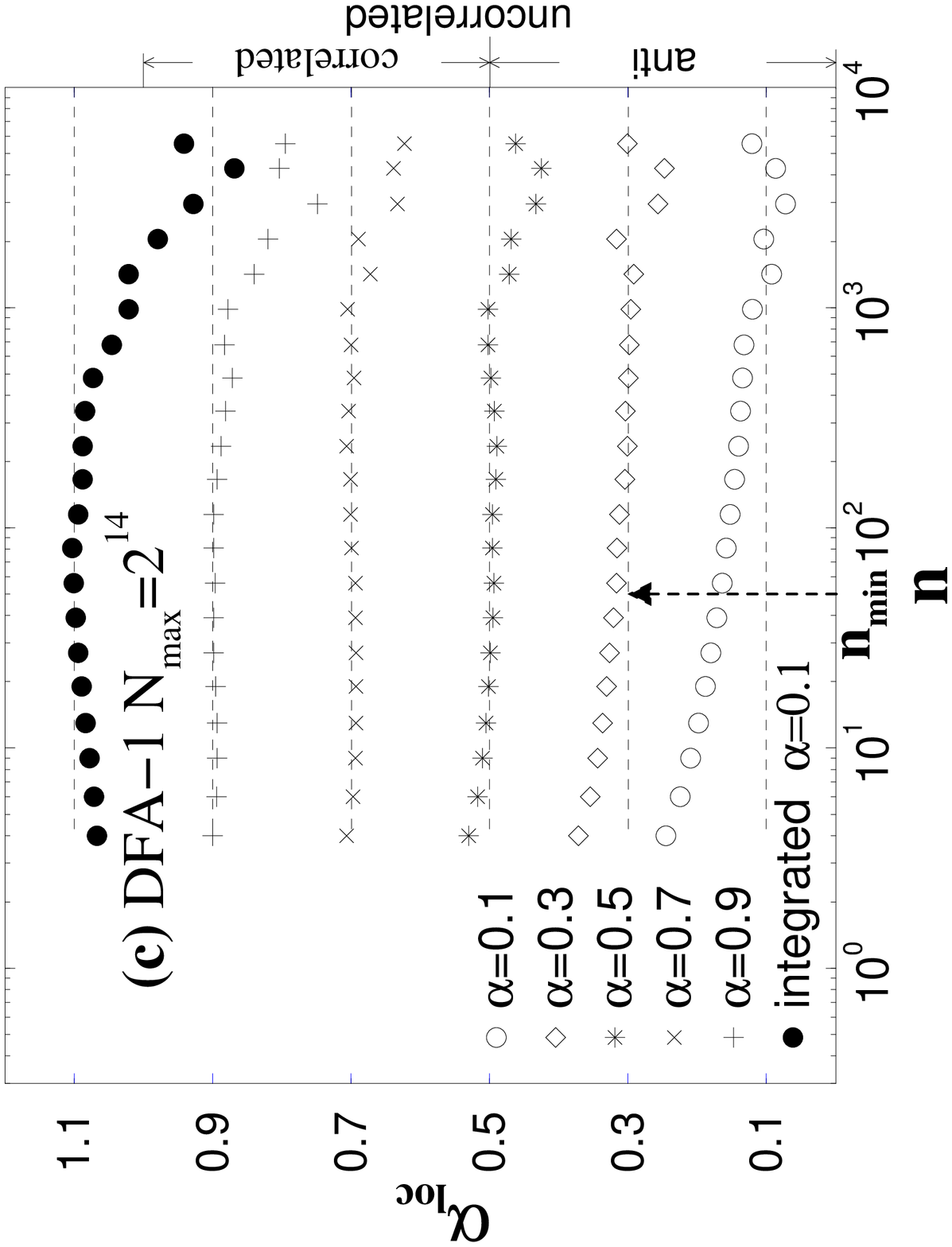}}}
}
\centerline{
\epsfysize=0.47\textwidth{\rotate[r]{\epsfbox{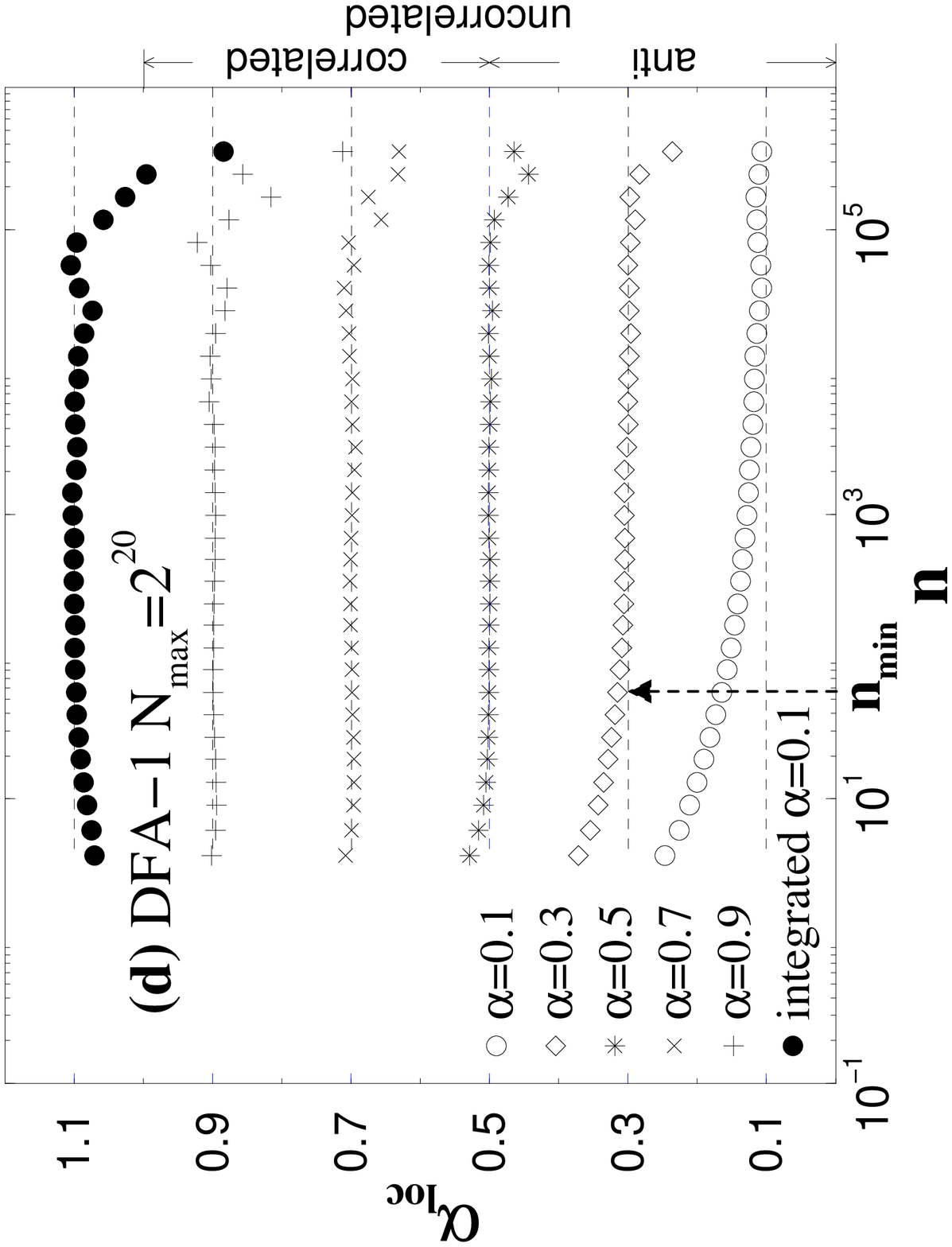}}}}
\caption{The estimated $\alpha$ from local fit (a) R/S analysis, the
length of signal $N_{\mbox{\scriptsize max}}=2^{14}$. (b)R/S analysis,
$N_{\mbox{\scriptsize max}}=2^{20}$. (c) DFA-1, $N_{\mbox{\scriptsize
max}}=2^{14}$ (d) DFA-1, $N_{\mbox{\scriptsize max}}=2^{20}$.
$\alpha_{\mbox{\scriptsize loc}}$ come from the average of $50$
simulations.  If a technique is working, then the data for scaling
exponent $\alpha$ should be a weakly fluctuating horizontal line
centered about $\alpha_{\mbox{\scriptsize loc}} = \alpha$. Note that
such a horizontal behavior does not hold for all the scales. Generally,
such a expected behavior begins from some scale $n_{\mbox{\scriptsize
min}}$, holds for a range and ends at a larger scale
$n_{\mbox{\scriptsize max}}$. For DFA-1, $n_{\mbox{\scriptsize min}}$ is
quite small $\alpha > 0.5$.  For R/S analysis, $n_{\mbox{\scriptsize
min}}$ is small only when $\alpha \approx 0.7$. } \label{slope_n}
\end{figure}

In order to find the best region, we first determine the dependence of
the locally estimated $\alpha$, $\alpha_{\mbox{\scriptsize loc}}$, on
the scale $n$. First, generate a standard noise
with given scaling exponent $\alpha$; then calculate $F(n)$ (or $R/S$),
and obtain $\alpha_{\mbox{\scriptsize loc}}(n)$ by local fitting of
$F(n)$ (or $R/S$). Same random simulation is repeated 50 times for both
DFA and R/S analysis.  The resultant average $\alpha_{\mbox{\scriptsize
loc}}(n)$, respectively, are illustrated in Fig.\ref{slope_n} for DFA-1
and R/S analysis.

If a scaling analysis method is working properly, then the result
$\alpha_{\mbox{\scriptsize loc}}(n)$ from simulation with $\alpha$ would
be a horizontal line with slight fluctuation centered about
$\alpha_{\mbox{\scriptsize loc}}(n) = \alpha$. Note from
Fig.\ref{slope_n} that such a \textit{horizontal behavior} does not hold
for all the scales $n$ but for a certain range from
$n_{\mbox{\scriptsize min}}$ to $n_{\mbox{\scriptsize max}}$. In
addition, at small scale, R/S analysis gives $\alpha_{\mbox{\scriptsize
loc}} > \alpha$ if $\alpha < 0.7$ and $\alpha_{\mbox{\scriptsize loc}} <
\alpha$ if $\alpha > 0.7$, which has been pointed out by
Mandelbrot\cite{estimateH} while DFA gives $\alpha_{\mbox{\scriptsize
loc}} > \alpha$ if $\alpha <1.0$ and $\alpha_{\mbox{\scriptsize loc}} <
\alpha$ if $\alpha >1.0$.

It is clear that the smaller the $n_{\mbox{\scriptsize min}}$ and the
larger the $n_{\mbox{\scriptsize max}}$, the better the method. We also
perceive that the expected \textit{horizontal behavior} stops because
the fluctuations become larger due to the under-sampling of $F(n)$ or
$R/S$ when $n$ gets closer to the length of the signal
$N_{\mbox{\scriptsize max}}$. Furthermore, it can be seen from
Fig.\ref{slope_n} that $n_{\mbox{\scriptsize max}} \approx
\frac{1}{10}N_{\mbox{\scriptsize max}}$ independent of $\alpha$ (if the
best fit region exists), which is why one tenth of the signal length is
the maximum box size when using DFA or R/S analysis.

\begin{figure}[H!]
\centerline{
\epsfysize=0.47\textwidth{\rotate[r]{\epsfbox{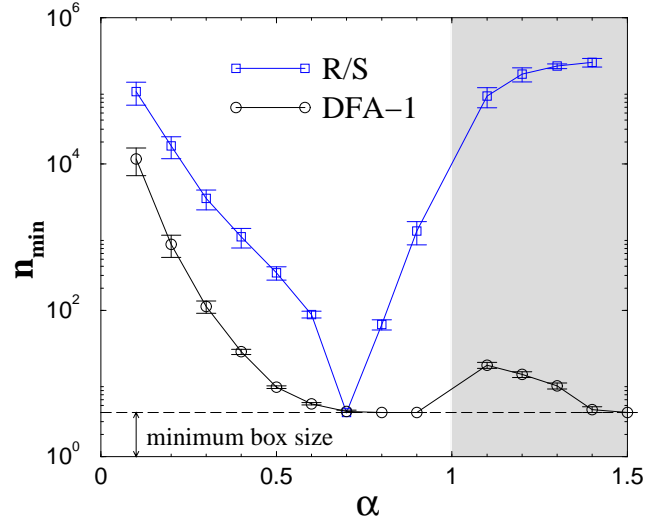}}}}
\vspace*{0.5cm}
\caption{The starting point of good fit region, $n_{\mbox{\scriptsize min}}$, for
DFA-1 and R/S analysis. The results are obtained from 50 simulations, in
which the length of noise is $N_{\mbox{\scriptsize max}}=2^{20}$. The
condition for a good fit is $\Delta \alpha = |\alpha_{\mbox{\scriptsize
loc}} - \alpha|< 0.01$. The data for $\alpha > 1.0$ shown in the shading
area are obtained by applying analysis on the integrations of noises
with $\alpha <1.0$. It is clear that DFA-1 works better than R/S analysis
because its $n_{\mbox{\scriptsize min}}$ is always smaller than that of
R/S analysis.  } \label{Smin}
\end{figure}
On the contrary, $n_{\mbox{\scriptsize min}}$ does not depend on the
$N_{\mbox{\scriptsize max}}$ since $\alpha_{\mbox{\scriptsize loc}}(n)$
at small $n$ hardly changes as $N_{\mbox{\scriptsize max}}$ varies but
it does depend on $\alpha$. Thus, we obtain $n_{\mbox{\scriptsize min}}$
quantitatively as shown in Fig.\ref{Smin}.  For R/S analysis, only for
$\alpha \approx 0.7$, $n_{\mbox{\scriptsize min}}$ is small; for
$\alpha$ a little away from $0.7$ (for example, 0.5),
$n_{\mbox{\scriptsize min}}$ becomes very large and close to
$n_{\mbox{\scriptsize max}}$, indicating that the best fit region will
vanish and R/S analysis does not work at all. Comparing to R/S, DFA
works better since $n_{\mbox{\scriptsize min}}$ is quite small for
$\alpha > 0.5$ correlated signals.

One problem remains for DFA, $n_{\mbox{\scriptsize min}}$ for small
$\alpha$ ($\leq 0.5$) is still too large comparing to those for large
$\alpha$($>0.5$). We can improve it by applying DFA on the integration
of the noise with $\alpha<0.5$. The resultant new expected $\alpha^{'}$
for the integrated signal would be $\alpha^{'}_0 = \alpha+1$, while the
$n_{\mbox{\scriptsize min}}$ for the integrated signal becomes much
smaller as shown also in Fig.\ref{Smin}(shading area $\alpha >
1$). Therefore, for a noise with $\alpha < 0.5$, it is best to
estimate the scaling exponent $\alpha^{'}$ of the integrated signal
first and then obtain $\alpha$ by $\alpha = \alpha^{'}-1$. This is what
we did in the following sections to those anticorrelated signals.

\section{Superposition law for DFA}\label{secadd}
For two uncorrelated signals $f(i)$ and $g(i)$, their root mean
square fluctuation functions are $F_f(n)$ and $F_g(n)$
respectively. We want to prove that for the signal
$f(i)+g(i)$, its rms fluctuation
\begin{equation}
F_{f+g}(n) = \sqrt{F_f(n)^2+F_g(n)^2}
\label{addlaw}
\end{equation}

Consider three signals in the same box first. The integrated signals for $f$, $g$ and $f+g$ are $y_f(i)$, $y_g(i)$ and $y_{f+g}(i)$ and their corresponding trends are $y^{fit}_{f}$, $y^{fit}_{g}$,$y^{fit}_{f+g}$ ($i=1,2,...,n$, $n$ is the box size).
Since $y_{f+g}(i)=y_{f}(i)+y_g(i)$ and combine the definition of detrended fluctuation function Eq.\ref{psi}, we have that for all boxes
\begin{equation}
Y_{f+g}(i)=Y_f(i)+Y_g(i),
\label{appaddpsi}
\end{equation}
where $Y_{f+g}$ is the detrended fluctuation function for the
signal $f+g$, $Y_f(i)$ is for the signal $f$ and $Y_g(i)$
for $g$. Furthermore, according to the definition of rms
fluctuation, we can obtain

\begin{eqnarray}
F_{f+g}(n) & & = \sqrt{\frac{1}{N_{max}} \sum\limits_{i=1}^{N_{max}}\left[Y_{f+g}(i)\right ]^2} \\\nonumber
& &=\sqrt{\frac{1}{N_{max}} \sum\limits_{i=1}^{N_{max}}\left[Y_{f}(i)+Y_{g}(i)\right ]^2},
\label{plugpsi}
\end{eqnarray}
where $\ell$ is the number of boxes and $k$ means the $k$\textit{th} box.
If $f$ and $g$ are not correlated, neither are $Y_{f}(i)$ and $Y_{g}(i)$ and, thus,
\begin{equation}
\sum\limits_{i=1}^{N_{max}}Y_{f}(i)Y_{g}(i)=0.
\label{uncorrelated}
\end{equation}
From Eq.\ref{uncorrelated} and Eq.\ref{plugpsi}, we have
\begin{eqnarray}
F_{f+g}(n) & & =\sqrt{\frac{1}{N_{max}} \sum\limits_{i=1}^{N_{max}}\left[Y_{f}(i)^2+Y_{g}(i)^2\right ]}\nonumber\\
& & =\sqrt{\left[F_{f}(n)\right ]^2+\left[F_{g}(n)\right ]^2}.
\end{eqnarray}

\section{DFA-1 on linear trend}\label{secdfa1l}
$\smallskip $Let us suppose a linear time series $u(i)=A_{\rm L}i$. The integrated signal $y_{L}(i)$ is
\begin{equation}
y_{L}(i)=\sum_{j=1}^{i}A_{\rm L}j=A_{\rm L}\allowbreak \frac{i^{2}+i}{2}
\end{equation}
Let as call $N_{max}$ the size of the series and $n$ the size of the box. The rms fluctuation $F_{\rm L}(n)$ as a function of $n$ and $N_{max}$ is

\end{multicols}
\renewcommand{\thesection}{\Alph{section}}
\begin{equation}
F_{\rm L}(n)=A_{\rm L}\sqrt{\frac{1}{N_{max}}\sum_{k=1}^{N_{max}/n} \sum_{i=(k-1)n+1}^{kn}\left(\frac{i^{2}+i}{2}-(a_{k}+b_{k}i)\right)^{2}}
\end{equation}

\begin{multicols}{2}

where $a_{k}$ and $b_{k}$ are the parameters of a least-squares fit of the $k
$-th box of size $n$. $a_{k}$ and $b_{k}$ can be determined analytically,
thus giving:
\begin{equation}
a_{k}=1-\frac{1}{12}n^{2}+\frac{1}{2}n^{2}k+\frac{1}{12}n-\frac{1}{2}%
k^{2}n^{2}
\end{equation}

\begin{equation}
b_{k}=1-\frac{1}{2}n+kn+\frac{1}{2}
\end{equation}
With these values, $F_{\rm L}(n)$ can be evaluated analytically:

\begin{equation}
F_{\rm L}(n)=A_{\rm L}\frac{1}{60}\sqrt{\left( 5n^{4}+25n^{3}+25n^{2}-25n-30\right) }
\label{lin}
\end{equation}
The dominating term inside the square root is $5n^{4}$ and then one obtains
\begin{equation}
F_{\rm L}(n)\approx \frac{\sqrt{5}}{60}A_{\rm L}n^{2}
\end{equation}
leading directly to an exponent of 2 in the DFA. An important consequence is that, as $F(n)$ does not depend on $N_{max}$, for linear trends
with the same slope, the DFA must give exactly the same results
for series of different sizes. This is not
true for other trends, where the exponent is 2, but the factor
multiplying $n^{2}$ can depend on $N_{max}$.

\section{DFA-1 on Quadratic trend}\label{secq}

Let us suppose now a series of the type $u(i)=A_{\rm Q} i^{2}$. The integrated time series $y(i)$ is

\begin{equation}
y(i)=A_{\rm Q}\sum_{j=1}^{i}j^{2}=A_{\rm Q}\frac{2i^{3}+3i^{2}+i}{6}
\end{equation}
As before, let us call $N_{max}$ and $n$ the sizes of the series
and box, respectively. The rms fluctuation function
$F_{\rm Q}(n)$ measuring the rms fluctuation is now defined as

\end{multicols}
\begin{equation}
F_{\rm Q}(n)=A_{\rm Q}\sqrt{\frac{1}{N_{max}}\sum_{k=1}^{N_{max}/n}%
\sum_{i=(k-1)n+1}^{kn}\left(\frac{2i^{3}+3i^{2}+i}{6}-(a_{k}+b_{k}i)\right)
^{2} }
\end{equation}
where $a_{k}$ and $b_{k}$ are the parameters of a least-squares fit of
the $k
$-th box of size $n$. As before, $a_{k}$ and $b_{k}$ can be determined
analytically, thus giving:
\begin{equation}
a_{k}=\frac{1}{15}n^{3}+n^{3}k^{2}-\frac{7}{15}n^{3}k+\frac{17}{30}n^{2
}k-%
\frac{7}{60}n^{2}+\frac{1}{20}n-\frac{2}{3}k^{3}n^{3}-\frac{1}{2}n^{2}k^{
2}+%
\frac{1}{15}kn
\end{equation}

\begin{equation}
b_{k}=\frac{3}{10}n^{2}+n^{2}k^{2}-n^{2}k+kn-\frac{2}{5}n+\frac{1}{10}
\end{equation}
Once $a_{k}$ and $b_{k}$ are known, $F(n)$ can be evaluated, giving:
\begin{equation}
F_{\rm Q}(n)=A_{\rm Q}\allowbreak \frac{1}{1260}\sqrt{-21\left(
n^{4}+5n^{3}+5n^{2}-5n-6\right) \left(
32n^{2}-6n-81-210N_{max}-140N_{max}^{2}\right) }
\end{equation}
\begin{multicols}{2}

As $N_{max}>n$, the dominant term
inside the square root is given by $140N_{max}^{2}\times 21n^{4}=A_{\rm Q}\allowbreak
2940n^{4}N_{max}^{2}$, and then one has approximately
\begin{equation}
F_{\rm Q}(n)\approx A_{\rm Q}
\frac{1}{1260}\sqrt{2940n^{4}N_{max}^{2}}=A_{\rm Q}\frac{1}{90}\sqrt{15}N_{max}n^{2}
\label{quad}
\end{equation}
leading directly to an exponent 2 in the DFA analysis. An interesting
consequence derived from Eq. (\ref{quad}) is that, $F_{\rm Q}(n)$ depends on the length of signal $N_{max}$, and the DFA line ($\log F_{\rm Q}(n)$ versus $\log n$) for
quadratic series $u(i)=A_{\rm Q} i^{2}$ of different $N_{max}$ DO NOT overlap (as it
happened for linear trends).


\end{multicols}

\end{document}